\documentclass[12pt]{article}
\usepackage{srcltx}
\usepackage{amsmath}
\usepackage{amsthm}
\usepackage{amssymb}
\usepackage{amsfonts}
\usepackage{epic}
\usepackage{eepic}
 \usepackage{times}
\usepackage[matrix,arrow]{xy}
\usepackage{macros}
\usepackage{epsfig}
\usepackage{bm}
\theoremstyle{plain}

\theoremstyle{remark}
\newtheorem{rem}{Remark}

\newcommand{\pa}{\partial}
\newcommand{\ot}{\otimes}
\newcommand{\ra}{\to}

\newcommand{\fr}[2]{{\textstyle \frac{#1}{#2} }}

\newcommand{\fsl}{{\mathfrak s}{\mathfrak l}}

\newcommand{\al}{\alpha}

\newcommand{\be}{\beta}
\newcommand{\ga}{\gamma}
\newcommand{\Ga}{\Gamma}
\newcommand{\de}{\delta}
\newcommand{\De}{\Delta}
\newcommand{\ep}{\epsilon}

\newcommand{\si}{\sigma}

\newcommand{\vf}{\varphi}

\newcommand{\CA}{{\mathcal A}}

\newcommand{\CC}{{\mathcal C}}
\newcommand{\CD}{{\mathcal D}}

\newcommand{\CF}{{\mathcal F}}
\newcommand{\CG}{{\mathcal G}}

\newcommand{\CM}{{\mathcal M}}
\newcommand{\CN}{{\mathcal N}}
\newcommand{\CO}{{\mathcal O}}

\newcommand{\CS}{{\mathcal S}}

\newcommand{\CV}{{\mathcal V}}

\newcommand{\SB}{{\mathsf B}}
\newcommand{\SC}{{\mathsf C}}
\newcommand{\SD}{{\mathsf D}}

\newcommand{\SF}{{\mathsf F}}

\newcommand{\sk}{{\mathsf k}}

\newcommand{\SM}{{\mathsf M}}
\newcommand{\SN}{{\mathsf N}}

\renewcommand{\SS}{{\mathsf S}}
\newcommand{\ST}{{\mathsf T}}
\newcommand{\SU}{{\mathsf U}}
\newcommand{\SV}{{\mathsf V}}
\newcommand{\SW}{{\mathsf W}}

\newcommand{\sm}{{\mathsf m}}

\newcommand{\sq}{{\mathsf q}}

\newcommand{\BR}{{\mathbb R}}

\newcommand{\BC}{{\mathbb C}}

\newcommand{\BP}{{\mathbb P}}

\newcommand{\BZ}{{\mathbb Z}}

\newcommand{\rf}[1]{(\ref{#1})}

\newcommand{\fus}[6]{F_{#5#6}^{}\big[\,{}^{#3}_{#4}\;{}^{#2}_{#1}\,\big]}
\newcommand{\mfus}[6]{F\big[\,{}^{#3}_{#4}\;{}^{#2}_{#1}\,\big]_{#5#6}^{}}
\newcommand{\gfus}[6]{G_{#5#6}^{}\big[\,{}^{#3}_{#4}\;{}^{#2}_{#1}\,\big]}

\newcommand{\nc}{\newcommand}

\nc{\rnc}{\renewcommand} \nc{\beq}{\begin{equation}}
\nc{\eeq}{\end{equation}} \nc{\beqa}{\begin{eqnarray}}
\nc{\eeqa}{\end{eqnarray}}

\begin{document}
\title{Isomonodromic tau-functions from Liouville conformal blocks}
\author{
N. Iorgov$\, ^a$\footnote{iorgov@bitp.kiev.ua}, 
O. Lisovyy$\,^b$\footnote{lisovyi@lmpt.univ-tours.fr}, 
J. Teschner$\,^c$\footnote{teschner@mail.desy.de}}
\address{
$^a$ Bogolyubov Institute for Theoretical Physics, 03680, Kyiv, Ukraine \\
$^b$ Laboratoire de Math\'ematiques et Physique Th\'eorique, Universit\'e de Tours, 37200 Tours, France \\ 
$^c$ DESY Theory, Notkestr. 85, 22603 Hamburg, Germany}
\maketitle

\begin{quote}{\small
The goal of this note is to show that the Riemann-Hilbert problem
to find multivalued analytic functions with ${\rm SL}(2,\BC)$-valued
monodromy on Riemann surfaces
of genus zero with $n$ punctures can be solved by taking suitable
linear combinations of the conformal blocks of Liouville theory at
$c=1$. This implies a similar representation for the isomonodromic tau-function.
In the case $n=4$ we thereby get a proof of the
relation between tau-functions and conformal blocks
discovered in \cite{GIL}. We briefly discuss a possible application of our results
to the study of relations between certain $\mathcal{N}=2$ supersymmetric gauge theories
and conformal field theory.
}
\end{quote}

\section{Introduction}

The problem to describe
isomonodromic deformations of ordinary differential equations has attracted
a lot of attention in the past. This is due to the existence of
a large number of
applications in various areas of mathematics and theoretical physics,
as well as the mathematical beauty and depth of the problem itself.

A first striking relation with quantum field theory was exhibited
in a series of papers of Sato, Miwa and Jimbo which appeared at the
end of the 1970's, see in particular \cite{SMJ}, and \cite{SMJ-R} for
a review.
The results include the identification of the
isomonodromic tau-functions, the generating functions for the Hamiltonians of the
isomonodromic flows,  with certain correlation functions
in a  quantum field theory of chiral free fermions.

The main result of this paper is another relation between conformal field  theory
and the isomonodromic deformation problem: The tau-functions for isomonodromic
deformations of flat ${\rm SL}(2)$-connections on $n$-punctured spheres
coincide with certain linear combinations of the Liouville conformal blocks at $c=1$.
This result leads in particular to a proof of the relation between Liouville conformal
blocks and the tau-function of Painlev\'e VI that was discovered in \cite{GIL}.

We are going to show that our result can be understood as a sort of
bosonization of the fermionic representations of tau-functions. To this aim we are
going to show that our construction is essentially equivalent to a bosonic construction of
the so-called twist fields whose insertion generates a singularity for the fermion field with specified
monodromy. In our approach the twist fields are constructed from the chiral vertex operators
of the Virasoro algebra.

Expressing the isomonodromic tau-functions in terms of Liouville conformal blocks
appears to have certain advantages compared to the previously known representations.
The famous formula for the asymptotics of Painlev\'e VI found by Jimbo \cite{Ji}, for example, is an easy
consequence. More generally, one may take advantage of the various results known
about the Liouville conformal blocks in order to get detailed information on the
isomonodromic tau-functions. Conversely, one may use this connection to find highly
non-trivial new results about the Liouville conformal blocks at $c=1$ \cite{ILT}.

As an interesting application
we are going to show
how the known algebro-geometric solutions of the Schlesinger system on $C_{0,n}$ \cite{kk} arise
from conformal blocks of the Ashkin-Teller critical model \cite{Za,ZZ}.

In the conclusions
we'll discuss a possible application of our results
to the study of $\mathcal{N}=2$ supersymmetric gauge theories: They
can be used to connect two recently discovered relations between certain classes of $\CN=2$,
$d=4$  supersymmetric gauge
theories on the one hand, and two-dimensional conformal field theories on the other hand.

The paper is organised as follows. In Section \ref{RHrev} we review the basic formulation of the
Riemann-Hilbert problem together with some basic material on the parameterization of monodromy
groups. The following Section \ref{CVO} collects the necessary background on Liouville conformal blocks.
Our main result is described in Section \ref{sec_solRHP}. We define infinite linear combinations of the
Virasoro conformal blocks, and show that the result solves the Riemann-Hilbert problem. Section 5 describes how to reformulate our results to
get a bosonic construction of twist fields creating singularities for
fermion fields with specified monodromy. The following Section 6 describes
two applications: We first rederive Jimbo's formula for the asymptotics of
Painlev\'e VI from our results, and show that specializing our construction
to Ashkin-Teller conformal blocks reproduces the algebro-geometric
solutions found in \cite{kk}. In the conclusions we indicate interesting
directions for future research including the application to
supersymmetric gauge theories mentioned above.

\section{The Riemann-Hilbert problem}\label{RHrev}

\subsection{Formulation of the Riemann Hilbert problem}

The fundamental group $\pi_1$
of $C_{0,n}:=\BP^1\setminus\{z_1,\dots,z_n\}$
has $n$ generators $\chi_1^{},\dots,\chi_n^{}$ subject to one relation
$\chi_1^{}\circ\chi_2^{}\circ\dots\circ\chi_n^{}=1$. 
Representations\footnote{Here understood as {\it anti}-homomorphisms
$\rho:\pi_1(C_{0,n})\ra{\rm SL}(2,\BC)$} $\rho$ of
$\pi_1(C_{0,n})$ in ${\rm SL}(2,\BC)$ are specified by collections
of matrices $M_k:=\rho(\chi_k^{})\in {\rm SL}(2,\BC)$, $k=1,\dots,n$
satisfying $M_n\cdot M_{n-1}\cdot \dots\cdot M_1=1$ up to overall
conjugation with elements of ${\rm SL}(2,\BC)$.
We will be interested in the cases where
the matrices $M_k$ are diagonalizable with
fixed eigenvalues $e^{\pm 2\pi i m_k}$. The space of all such representations
of $\pi_1(C_{0,n})$ is then $2(n-3)$-dimensional.

It will be convenient to choose a base-point $y_0$ on $C_{0,n}$.
The dependence on the choice of $y_0$ will turn out to be inessential.
We may then represent the generators $\chi_k$ by closed paths starting and
ending at $y_0$. The Riemann-Hilbert problem is to find a multivalued
analytic
matrix
function $Y(y)$ on $C_{0,n}$ such that the monodromy along
$\chi_k$ is represented as
\begin{equation}\label{monoact}
Y(\chi_k^{}.y)\,=\,Y(y)M_k\,,
\end{equation}
where $Y(\chi_k^{}.y)$ denotes the analytic continuation of
$Y(y)$ along $\chi_k$.

The solution to this problem is unique up to left multiplication with
single valued matrix functions. In order to fix this ambiguity we need to
specify the singular behavior of $Y(y)$, leading to the following
refined version of the Riemann-Hilbert problem: Find a matrix
function $Y(y)$ such that the following conditions are satisfied.
\begin{itemize}
\item[i)] $Y(y_0)\,=\,1\,$,
\item[ii)] $Y(y)$ is a multivalued, analytic and invertible
on $C_{0,n}$,
\item[iii)] There exist neighborhoods of $z_k$, $k=1,\dots,n$ where
$Y(y)$ can be represented as
\begin{equation}\label{asym}
Y(y)\,=\,\hat{Y}^{(k)}(y)\,(y-z_k)^{\mu_k}\,,\qquad
M_k=e^{2\pi{\mathrm i}\mu_k}\,,
\end{equation}
with $\hat{Y}^{(k)}(y)$ being holomorphic and invertible at $y=z_k$
and $\mu_1,\dots,\mu_n\in\fsl(2,\BC)$.
\end{itemize}
If such function $Y(y)$ exists, it is uniquely determined by the
monodromy data $\mu=(\mu_1,\dots,\mu_n)$.

The refined Riemann-Hilbert problem naturally arises in the study
of rank 2 flat connections on $C_{0,n}$. Any flat connection
on $C_{0,n}$ is gauge equivalent to a holomorphic connection of the
form $\pa_y-A(y)$,  with $A(y)$ of the form
\begin{equation}
A(y)\,=\,\sum_{k=1}^n \frac{A_k}{y-z_k}\,,
\end{equation}
where $A_1,\dots A_n\in \fsl(2,\BC)$, $\sum_{k=1}^nA_k=0$.
One may then consider the fundamental
matrix solution
$Y(y)$ of the differential equation
\begin{equation}
\frac{\pa}{\pa y}Y(y)\,=\,A(y)Y(y)\,,
\end{equation}
normalized
by $Y(y_0)=1$. It will automatically satisfy ii) and iii) for certain
$\mu_1,\dots,\mu_n$, provided that the eigenvalues $\pm m_k$ of
$A_k$ satisfy the
condition $2m_k\notin\BZ$. Any representation $\rho:\pi_1(C_{0,n})\ra
{\rm SL}(2,\BC)$ can be realized as monodromy representation of such a
Fuchsian system, which means that a solution to the Riemann-Hilbert problem
formulated will generically exist.
The Riemann-Hilbert correspondence between flat connections $\pa_y-A(y)$ and
representations $\rho:\pi_1(C_{0,n})\ra {\rm SL}(2,\BC)$ allows us
to identify the moduli space $\CM_{\rm flat}(C_{0,n})$ of flat
$\fsl(2,\BC)$-connections on $C_{0,n}$ with the so-called character
variety
${\rm Hom}(\pi_1(C_{0,n}),{\rm SL}(2,\BC))/{\rm SL}(2,\BC)$.

\subsection{Trace coordinates}\label{Darboux}

Useful sets of coordinates for
$\CM_{\rm flat}(C_{0,n})$ are given by the trace functions
$L_{\ga}:=\operatorname{\rm tr}\rho(\ga)$ associated to any
simple closed curve $\ga$ on $C_{0,n}$.
Minimal sets of trace functions that can be used to
parameterize $\CM_{\rm flat}(C_{0,n})$ can be identified using
pants decompositions.
In order to have uniform
notations, let us replace the punctures $z_1,\dots,z_n$ by little holes
obtained by cutting along non-intersecting simple closed curves $\de_k$
surrounding the punctures $z_k$, $k=1,\dots,n$, respectively.
A pants decomposition is defined by cutting $C_{0,n}$ along $n-3$
simple closed curves $\ga_r$, $r=1,\dots,n-3$ on $C_{0,n}$.
This will
decompose $C_{0,n}$ into a disjoint union of
$n-2$ three-holed spheres $C_{0,3}^t$, $t=1,\dots,n-2$.
The collection $\CC=\{\ga_1,\dots,\ga_{n-3}\}$ of curves will be
called the cut system.

To each curve $\ga_r\in\CC$
let us associate the union of the
two three-holed spheres that have $\ga_r$
in its boundary, a four-holed sphere $C_{0,4}^r$.
It will be assumed that the
curves $\ga_r$, $r=1,\dots,n-3$
are oriented.
The orientation of $\ga_r$ allows us to introduce a natural
numbering of the boundaries of $C_{0,4}^r$. We may then
consider the curves $\ga^r_s$ and $\ga_t^r$ which encircle
the pairs of
boundary components of $C_{0,4}^r$ with numbers $(1,2)$ and
$(2,3)$, respectively. The corresponding trace functions
will be denoted as $L^r_s$ and $L_t^r$. The collection
of pairs of trace functions $(L^r_s,L_t^r)$, $r=1,\dots,n-3$
can be used to parameterize $\CM_{\rm flat}(C_{0,n})$.

A closely related set of coordinates for
$\CM_{\rm flat}(C_{0,n})$ is obtained by
parameterizing $L^r_s$ and $L_t^r$ in terms of
complex numbers
$(\si_r,\tau_r)$ as
\begin{subequations}\label{classWT}
\begin{align}
&L^r_s\,=\,2\cos 2\pi \si_r\,,\label{Wilson}\\
 (\sin (2\pi  \si_r))^2\,&L^r_t\,=\,
C_+(\si_r)\,e^{i\tau_r}+C_0(\si_r)+C_-(\si_r)\,e^{-i\tau_r}\,,
\end{align}
\end{subequations}
where
\begin{subequations}\label{Cepdef}
\begin{align}
C_{+}(\si_r)&=-4\prod_{s=\pm 1} \sin\pi(\si_r+s(\si^r_1-\si^r_2))\sin\pi(\si_r+s(\si^r_3-\si^r_4))\,,\\
C_0(\si_r)&={2 }\,
\big[\cos 2\pi  \si^r_2 \cos 2\pi \si^r_3 + \cos 2\pi  \si^r_1 \cos 2\pi \si^r_4\big]\\
&\quad- {2 \cos 2\pi  \si_r}
\big[\cos 2\pi  \si^r_1 \cos 2\pi \si^r_3 + \cos 2\pi  \si^r_2 \cos 2\pi \si^r_4\big]
\, ,\notag\\
C_{-}(\si_r)&=-4\prod_{s=\pm 1} \sin\pi(\si_r+s(\si^r_1+\si^r_2))\sin\pi(\si_r+s(\si^r_3+\si^r_4))\,.
\end{align}
\end{subequations}
In order to define $\si^r_i$, $i=1,\dots,4$ in \rf{Cepdef}
let us note that the boundary of $C_{0,4}^r$ with label $i$ may
either be a curve $\ga_{r'}\in\CC$, or it must coincide with a
curve $\de_k$ surrounding puncture $z_k$. We will identify
$\si^r_i\equiv\si_{r'}$ in the first case, while $\si^r_i$ will be identified
with an eigenvalue of $\mu_k$ otherwise.

The collection of data $(\si_r,\tau_r)$, $r=1,\dots,n-3$ will be denoted
as $(\si,\tau)$.
We observe that the coordinates $(\si,\tau)$ are for $n=4$
close relatives of the parameters used in \cite{Ji}.
They are also closely related to the coordinates used
in \cite{NRS}.

\subsection{Isomonodromic deformations and tau-function}

Let us briefly recall the well-known
relations to the isomonodromic deformation problem. Given a solution $Y(y)$
to the Riemann-Hilbert problem we may define an associated connection $A(y)$
as
\begin{equation}
A(y)\equiv A(y|z):=\,(\pa_y Y(y))\cdot (Y(y))^{-1}\,,
\end{equation}
It follows from \rf{asym} that
\begin{equation}
A(y|z)\,=\,\sum_{k=1}^{n}\frac{A_k(z)}{y-z_k}\,.
\end{equation}
It is well-known that variations of the positions $z_r$ will not change the
monodromies of the connection $A(y)$ provided that the
matrix residues $A_k=A_k(z)$ satisfy the following equations,
\begin{equation}
\begin{aligned}
&\pa_{z_k}A_k\,=\,-\sum_{l\neq k}
\frac{[A_k,A_l]}{z_k-z_l}\,,\\
&\pa_{z_l}A_k\,=\,\frac{y_0-z_k}{y_0-z_l}\frac{[A_k,A_l]}{z_k-z_l}\,,\quad
k\neq l\,,
\end{aligned}
\qquad \pa_{y_0}A_k\,=\,-\sum_{l\neq k}\frac{[A_l,A_k]}{y_0-z_l}\,.
\end{equation}
In the limit $y_0\ra\infty$ one finds the Schlesinger equations
\begin{equation}\label{Schlesinger}
\begin{aligned}
&\pa_{z_k}A_k\,=\,-\sum_{l\neq k}
\frac{[A_k,A_l]}{z_k-z_l}\,,\\
&\pa_{z_l}A_k\,=\,\frac{[A_k,A_l]}{z_k-z_l}\,,\quad
k\neq l\,.
\end{aligned}
\end{equation}
The
Schlesinger equations define Hamiltonian flows, generated by
the Hamiltonians
\begin{equation}
H_k:=\frac{1}{2}\operatorname{res}_{y=z_k}\operatorname{tr}A^2(y)=\sum_{l\neq k}
\frac{{\rm tr}(A_kA_l)}{z_l-z_k}\,,
\end{equation}
using the Poisson structure
\begin{equation}
\big\{\,A\left(y\right)\,\substack{\otimes\vspace{-0.1cm} \\ ,}\,A\left(y'\right)\,\big\}\,=\,
 \left[\,\frac{\mathcal{P}}{y-y'}\,,\,A\left(y\right)\otimes 1+1\otimes A\left(y'\right)\,\right],
\end{equation}
where $\mathcal{P}$ denotes the permutation matrix.
The tau-function $\tau(z)$ is defined as the generating function for the
Hamiltonians $H_k$,
\begin{equation}\label{taudef}
H_k\,=\, \pa_{z_k}\log\tau(z)\,.
\end{equation}
Integrability of \rf{taudef} is ensured by the Schlesinger equations
\rf{Schlesinger}.

\section{Chiral vertex operators and conformal blocks}\label{CVO}

Let us introduce the necessary definitions and results on the representation
theory of the Virasoro algebra which has generators $L_n$, $n\in\BZ$
and relations
\begin{equation}\label{Vir}
[\,L_n\,,\,L_m\,]\,=\,(n-m)L_{n+m}+\frac{c}{12}n(n^2-1)\de_{n+m,0}\,.
\end{equation}
Although we will ultimately be interested in the case $c=1$, it
will be useful to consider more general values of $c$ in some of
our arguments.
Highest weight representations $\CV_{\al}$ are generated from
vectors $|\al\rangle$ which satisfy
\begin{equation}
L_n\,|\,\al\,\rangle\,=\,0\,,\quad n>0\,,\quad
L_0\,|\,\al\,\rangle\,=\,\De_\al\,
|\,\al\,\rangle\,,
\end{equation}
where $\De_\al=\al(Q-\al)$ if $c$ is parameterized as
$c=1+6Q^2$. The representations  $\CV_{\al}$ can be decomposed into
the so-called energy-eigenspaces
\begin{equation}
\CV_{\al}\,\simeq\,\bigoplus_{n\in\BZ_{\geq 0}}\CV_{\al}^{(n)}\,,
\end{equation}
defined by the condition
$L_0v=(\De_{\al}+n)v$ for all $v\in \CV_{\al}^{(n)}$.

\subsection{Chiral vertex operators}

Chiral vertex operators $V^{\al}_{\be_2\be_1}(z)$ can be defined as
operators that map $\CV_{\be_1}\ra\CV_{\be_2}$ such that
\begin{equation}\label{VOdef}
L_n\,V^{\al}_{\be_2\be_1}(z)-V^{\al}_{\be_2\be_1}(z)\,L_n\,=\,
z^n(z\pa_z+\De_{\al}(n+1))V^{\al}_{\be_2\be_1}(z)\,.
\end{equation}
We have in particular
\begin{equation}\label{normcond}
V^{\al}_{\be_2\be_1}(z)\,|\,\be_1\,\rangle\,=\,
N(\be_2,\al,\be_1)\,z^{\De_{\be_2}-\De_{\be_1}-\De_\al}\,
\big[\, |\,\be_2\,\rangle+\CO(z)\,\big]\,,
\end{equation}
with a normalization factor $N(\be_2,\al,\be_1)$ that will be specified
later. It is well-known that the conditions (\ref{VOdef}) define
$z^{\De_{\be_1}+\De_\al-\De_{\be_2}}V^{\al}_{\be_2\be_1}(z)$
uniquely in the sense of formal power series in~$z$,
\begin{equation}\label{V:z-exp}
V^{\al}_{\be_2\be_1}(z)\,=\,z^{\De_{\be_2}-\De_{\be_1}-\De_\al}
\sum_{n=-\infty}^{\infty}z^n \,W^{\al}_{\be_2\be_1}(n)\,,\qquad
W^{\al}_{\be_2\be_1}(n):\CV_{\be_1}^{(k)}\ra \CV_{\be_2}^{(k+n)}\,.
\end{equation}
It has furthermore been argued in \cite{T03}
that the composition $V^{\al_2}_{\be_3\be_2}(z)V^{\al_1}_{\be_2\be_1}(w)$
of such vertex operators exists for $|w/z|<1$,
and that matrix elements such as
\begin{equation}
\langle \,\al_n\,|\,V_{\al_n\be_{n-3}}^{\al_{n-1}}(z_{n-1})
V_{\be_{n-3}\be_{n-4}}^{\al_{n-2}}(z_{n-2})\cdots
V_{\be_{1}\al_1}^{\al_2}(z_{2})\,
|\,\al_1\,\rangle\,,
\end{equation}
are represented by absolutely convergent power series in
$z_{k}/z_{k+1}$, $k=2,\dots,n-2$.

From each chiral vertex operator $V^{\al}_{\be_2\be_1}(z)$
one may generate a family of vertex operators called
descendants of $V^{\al}_{\be_2\be_1}(z)$. The descendants
of  $V^{\al}_{\be_2\be_1}(z)$ are labelled by the vectors in $\CV_{\al}$,
and the descendant corresponding to $v\in\CV_{\al}$
will be denoted as $V^{\al}_{\be_2\be_1}[v](z)$. The descendants
may be defined by means of the recursion relations
\begin{subequations}\label{descdef}\begin{align}
& V^{\al}_{\be_2\be_1}[|\al\rangle](z)\,\equiv\,V^{\al}_{\be_2\be_1}(z)\,,\\
& V^{\al}_{\be_2\be_1}[L_{-1}v](z)\,\equiv\,\pa_zV^{\al}_{\be_2\be_1}[v](z)\,,\\
& V^{\al}_{\be_2\be_1}[L_{-2}v](z)\,\equiv\,:T(z)V^{\al}_{\be_2\be_1}[v](z):\,,
\label{descdefL-2}\end{align}
\end{subequations}
where the following notation  has been used in \rf{descdefL-2}:
\begin{equation}\label{normord}
:T(z)V^{\al}_{\be_2\be_1}[v](z):\equiv \sum_{k<-1}z^{-k-2}L_k \,V^{\al}_{\be_2\be_1}[v](z)+\sum_{k\geq-1}z^{-k-2} \,V^{\al}_{\be_2\be_1}[v](z)\,L_k\,.
\end{equation}
The recursion relations \rf{descdef} suffice to define $V^{\al}_{\be_2\be_1}[L_{-n}v](z)$ for all $n>0$ thanks to the Virasoro algebra \rf{Vir}.

Using the descendants one may define a trilinear form $\CC_{0,3}:\CV_{\al_3}\ot
\CV_{\al_2}\ot\CV_{\al_1}\ra \BC$ as
\begin{equation}
\CC_{0,3}(v_3\ot v_2\ot v_1):=\langle \,v_3\,|\,
V^{\al_2}_{\al_3\al_1}[v_2](z)\,
|\, v_1\,\rangle\,.
\end{equation}
This trilinear form can be identified with the conformal block associated to
the three-punctured sphere $C_{0,3}$.

The definition of descendants allows us to introduce another way to compose
chiral vertex operators. We may e.g. consider
\begin{equation}
V^{\be_3}_{\be_2\be_1}\big[\,V^{\al_2}_{\be_3\al_1}[v_2](w-z)v_1\,\big](z)\,,
\end{equation}
which is defined a priori as a formal power series in $w-z$. Quadrilinear forms
such as
\begin{equation}
\CC_{0,4}(v_4\ot \dots \ot v_1):=\langle \,v_4\,|\,
V^{\be}_{\al_4\al_1}\big[\,V^{\al_3}_{\be\al_2}[v_3](w-z)v_2\,\big](z)\,
|\, v_1\,\rangle\,,
\end{equation}
will define absolutely convergent series in $w-z$ for all $v_4,\dots,v_1$
of finite energy. The quadrilinear forms $\CC_{0,4}(v_4\ot \dots \ot v_1)$
can be identified with conformal blocks associated to
the four-punctured sphere $C_{0,4}$.

By using the two types of composition of chiral vertex operators introduced
above one may construct conformal blocks associated to arbitrary
pants decompositions of $n$-punctured spheres.

\subsection{Degenerate fields}

Of particular importance for us will be the special case where $\al=-b/2$,
assuming that $Q$ is represented as $Q=b+b^{-1}$.
If furthermore $\be_2$ and $\be_1$ are related as
$\be_2=\be_1\mp b/2$, the vertex operators
$\psi_s(y)\equiv\psi_{\be_1,s}(y):=V^{-b/2}_{\be_1- sb/2,\be_1}(y)$, $s=\pm 1$,
are well-known to satisfy a differential equation
of the form
\begin{equation}\label{null}
\pa_y^2\psi_{\be_1,s}(y)+b^2:T(y)\psi_{\be_1,s}(y):=\,0\,,
\end{equation}
with normal ordering defined in \rf{normord}.
The chiral vertex operators
$\psi_{\be_1,s}(y)$ are called degenerate
fields. It follows from \rf{null} that matrix elements such as
\begin{align}\label{CFdef}
&\CF(\,\al;\be\,|\,z\,|\,y_0\,|\,y\,):=
\langle \,\al_n\,|\,\psi_{s'}(y_0) \psi_{s}(y) \,|\,\Theta\,
\rangle\,,\\
&|\,\Theta\,
\rangle:=
V_{\al_n+(s+s')\frac{b}{2},\be_{n-3}}^{\al_{n-1}}(z_{n-1})
V_{\be_{n-3}\be_{n-4}}^{\al_{n-2}}(z_{n-2})\cdots
V_{\be_{1}\al_1}^{\al_2}(z_{2})V_{\al_1,0}^{\al_1}(z_1)
|\,0\,\rangle\,,
\notag\end{align}
will satisfy the partial differential equation $\CD_{\rm BPZ} \CF=0$,
with
\begin{equation}\label{BPZ}
\CD_{\rm BPZ}:=\, \frac{1}{b^2}\frac{\pa^2}{\pa y^2}+
\frac{\De_{-\frac{b}{2}}}{(y-y_0)^2}+\frac{1}{y-y_0}\frac{\pa}
{\pa y_0}+
\sum_{k=1}^{n-1}\left(\frac{\De_{\al_k}}{(y-z_k)^2}+\frac{1}{y-z_k}\frac{\pa}
{\pa z_k}\right)\,,
\end{equation}
together with a similar differential equation for $y_0$.
Using this differential equation it may be shown that
$\CF(\,\al;\be\,|\,z\,|\,y_0\,|\,y\,)$, considered as a function
of $y$, can be analytically continued to a
multivalued analytic function on $C_{0,n}$.

\subsection{Braiding and fusion of degenerate fields}\label{fusbraid}

\begin{figure}[t]
\epsfxsize6.5cm
\centerline{\epsfbox{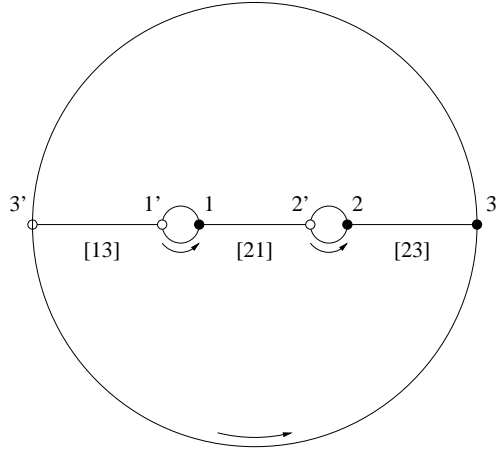}}
\caption{\it A sphere with three holes.
The arrows indicate our orientation conventions.}
\label{pantsdef}\vspace{.3cm}
\end{figure}

The differential equations \rf{null} satisfied by the degenerate fields
can be used to get a precise description of the monodromies of
the conformal blocks $\CF(\,\al;\be\,|\,z\,|\,y_0\,|\,y\,)$ defined in
\rf{CFdef}. Let us briefly summarize the relevant results.
There are three ways to compose a degenerate field with a
generic chiral vertex operator,
\begin{equation}\label{compo}
\begin{aligned}
& (1) \quad V_{\al_3,\al_1-s\frac{b}{2}}^{\al_2}[v_2](z)\,\psi_{s}(y)\,,\\
& (3) \quad \psi_{-s}(y)V_{\al_3-s\frac{b}{2},\al_1}^{\al_2}[v_2](z)\,,
\end{aligned}\qquad
(2)\quad
V_{\al_{3},\al_1}^{\al_2-s\frac{b}{2}}
\big[\psi_{s}(y-z) v_2\big](z)\,.
\end{equation}
The three ways \rf{compo} to compose these vertex operators correspond to
having the degenerate field $\psi_s(y)$ located
in the vicinity of the boundary components
with labels 1, 2 and 3, respectively, referring to Figure \ref{pantsdef} for
the notations.
The conformal blocks defined using the
three compositions \rf{compo}
are single valued and analytic in neighborhoods of the
black dots marked in Figure \ref{pantsdef} on the boundaries of the
three holes of $C_{0,3}$, respectively.
We are going to describe their
analytic continuation to the universal cover of $C_{0,3}$.
It will be helpful to introduce a
separate notation for
the vertex operator $\psi_{s}(y)$ when it is inserted at the antipodal
point of the circle $|y|={\rm const}$,
\begin{equation}\label{halfmono}
\psi_{\be,s}'(y)\,=\,B_s(\be)\psi_{\be,s}(e^{-\pi i}y)\,,\qquad
B_{s}(\al)\,=\,e^{\pi i(
\De_{\al-s\frac{b}{2}}-\De_{\al}-\De_{-\frac{b}{2}})}\,.
\end{equation}
The vertex operators $\psi_{\be,s}'(y)$ are single-valued in
an open neighborhood containing segments of the negative real axis.
One may naturally consider compositions (1)'-(3)'
of the form \rf{compo}, but with $\psi_s(y)$ replaced by
$\psi_s'(y)$. Regions on $C_{0,3}$
where the compositions (1)'-(3)' define single-valued
analytic conformal blocks are neighbourhoods of the
small empty circles in Figure \ref{pantsdef}.

The main building block for the monodromies will be the
following relations,
\begin{subequations}\label{moves}
\begin{align} \label{fus}
\psi_{-s_1}(y)V_{\al_3-s_1\frac{b}{2},\al_1}^{\al_2}[v_2](z)
& =\sum_{s_2=\pm 1} F_{s_1,s_2}^{[23]}\,
V_{\al_{3},\al_1}^{\al_2-s_2\frac{b}{2}}
\big[\psi_{s_2}(y-z) v_2\big](z),
\\
V_{\al_3,\al_1-s_1\frac{b}{2}}^{\al_2}[v_2](z)\,\psi_{s_1}(y)
& =\sum_{s_2=\pm 1} F_{s_1,s_2}^{[21]}\,
V_{\al_{3},\al_1}^{\al_2-s_2\frac{b}{2}}
\big[\psi_{s_2}'(y-z) v_2\big](z)\,,
\\
\psi_{-s_1}'(y)\,V_{\al_3-s_1\frac{b}{2},\al_1}^{\al_2}[v_2](z)
 & =\sum_{s_2=\pm 1}
F_{s_1,s_2}^{[13]}\,V_{\al_3,\al_1-s_2\frac{b}{2}}^{\al_2}[v_2](z)\,
\psi_{s_2}'(y)
\label{braid}
\,.
\end{align}
\end{subequations}
The relevant transport matrices are given respectively as
\begin{align}
F_{s_1,s_2}^{[ji]}
=\frac{\Ga(1+s_1b(2\al_i-Q))\Ga(s_2b(Q-2\al_j))}
{\prod_{s_3=\pm}
\Ga\big(\frac{1}{2}+s_1b(\al_i-Q/2)-s_2b(\al_j-Q/2)+s_3 b(\al_k-Q/2)\big)}\,,
\end{align}
valid if the vertex operators $V^{\al}_{\be_2,\be_1}(z)$ are normalized
via \rf{normcond} with $N(\al_3,\al_2,\al_1)\equiv 1$.

\begin{rem}
Comparing with the Moore-Seiberg formalism let us note that
\begin{subequations}
\begin{align}
F_{s_1,s_2}^{[21]}\,\equiv\,\mfus{\;\;\;{\al}_1}{-b/2}{\al_2}{\al_3}{s_1}{s_2}\,\equiv\,
\fus{\;\;\;{\al}_1}{-b/2}{\al_2}{\al_3}{\al_1-s_1\frac{b}{2};}{\al_2-s_2\frac{b}{2}}\,,\\
F_{s_1,s_2}^{[23]}\,\equiv\,\mfus{{\al}_1}{\al_2}{-b/2}{\;\;\;\al_3}{s_1}{s_2}\,\equiv\,
\fus{{\al}_1}{\al_2}{-b/2}{\;\;\;\al_3}{\al_3-s_1\frac{b}{2};}{\al_2-s_2\frac{b}{2}}\,,\\
F_{s_1,s_2}^{[13]}\,\equiv\,\mfus{{\al}_1}{\al_2}{\;\;\;\al_3}{-b/2}{s_1}{s_2}\,\equiv\,
\fus{{\al}_1}{\al_2}{\;\;\;\al_3}{-b/2}{\al_3-s_1\frac{b}{2};}{\al_1-s_2\frac{b}{2}}\,,
\end{align}
\end{subequations}
The relevant fusion matrices are related to each other
by the symmetries
\begin{equation}
\mfus{\;\;\;{\al}_1}{-b/2}{\al_2}{\al_3}{s_1}{s_2}=
\mfus{\al_3}{\al_2}{-b/2}{\;\;\;{\al}_1}{s_1}{s_2}
=\mfus{-b/2}{\;\;\;{\al}_1}{\al_3}{\al_2}{s_1}{s_2}
\,,
\end{equation}
together with
\begin{equation}
F\big[\,{}_{\al_3}^{\al_2} \;{}_{\;\;\;\al_1}^{-b/2}\,\big]^{-1}
=F\big[\,{}_{\al_3}^{\al_1}\;{}_{\;\;\;\al_2}^{-b/2}\,\big]\,.
\end{equation}
The definition of the
antipodal vertex operators $\psi_{\be,s}'(y)$ in \rf{halfmono}
is related to the elementary braid relation
\begin{equation}\label{elembraid}
[\,V_{\al_3,\al_1}^{\al_2}(y)\, V_{\al_1,0}^{\al_1}(z) |\,0\,\rangle\,]_{\circlearrowleft}^{}\,=\,
\Omega^{\al_3}_{\al_2,\al_1}\,V_{\al_3,\al_2}^{\al_1}(z)\,
V_{\al_2,0}^{\al_2}(y) |\,0\,\rangle\,,
\end{equation}
with left hand side defined by means of analytic continuation making
$y$ encircle $z$ in the anti-clockwise sense.
It is easy to see that the ``half-monodromy'' used in \rf{halfmono}
is related to the composition of analytic continuation \rf{elembraid} with
a suitable translation. It follows that the braiding phase factor
$B_s(\al)$
is related to the factors $\Omega^{\al_3}_{\al_2,\al_1}$
in \rf{elembraid} as
$B_s^{}(\al)\,=\,\Omega^{\al-sb/2}_{-b/2,\al}\,.
$
\end{rem}

In the normalisation where $N(\al_3,\al_2,\al_1)\equiv 1$
one may observe that the conformal blocks and the
fusion matrices $F^{[ji]}$
are perfectly analytic with
respect to the central charge $c$. We may in particular take the
limit $c\ra 1$ without encountering any problem. This is not the case for the kernel
of the integral transformation relating conformal blocks associated to different pants
decompositions.

\subsection{Monodromy action on spaces of conformal blocks}\label{q-mono}

Using these ingredients it is straightforward to show that the
analytic continuation of the matrix elements
$\CF(\,\al;\be\,|\,z\,|\,y_0\,|\,y\,)$ along the closed
paths $\chi_k$ can be expressed as a linear
combination of the matrix elements
$\CF(\,\al;\be'\,|\,z\,|\,y_0\,|\,y\,)$
having parameters $\be'_r$ that differ from $\be_r$ by integer
multiples
of the parameter $b$. In order to have a convenient notation let us
define the shift operators $\SV_{\be_r}^{\frac{1}{2}}$ which acts on functions
{\it to the left} as
\begin{equation}\label{shiftdef}
\CF(\,\al;\be\,|\,z\,|\,y_0\,|\,y\,)\cdot \SV_{\be_r}^{\frac{1}{2}}\,=\,
\CF(\,\al;\be-\fr{b}{2}e_r\,|\,z\,|\,y_0\,|\,y\,)\,,
\end{equation}
where $e_r$ is the vector in $\BC^{n-3}$ with components $\de_{rs}$.

\subsubsection{Geometrical set-up} \label{param}

It will be useful for
us to refine the pants decompositions as follows.
On each curve $\ga$ in the extended
cut system $\hat\CC:=\{\ga_1,\dots,\ga_{2n-3}\}$, where
$\ga_{n-3+k}:=\de_k$ for $k=1,\dots,n$ let us mark two points, a black one
and a white one.
On each pair of pants with label $t$ let us introduce a collection of two
non-intersecting arcs  $[23]_t$,  and
$[13]_t$ that connect marked points on the
boundary components labelled by $1$, $2$ and $3$, respectively.
These contours are depicted in Figure \ref{pantsdef}.



Let us next note that
any generator $\chi_k^{}$ of $\pi_1(C_{0,n})$ may be represented
as a concatenation $\eta_1\circ \eta_2
\circ\dots\circ \eta_N$ of oriented
arcs $\eta_a$, each contained within a
three-holed sphere $C_{0,3}^t$. 
It will not cause a loss
of generality to assume that each arc $\eta_a$ is
of the following two types:
\begin{itemize}
\item An arc $[ji]_t$ on $C_{0,3}^t$
running from the marked point on
boundary component $i$ of trinion (three punctured sphere) $t$
to the one on boundary component $j$ as depicted in
Figure \ref{pantsdef},
\item An arc $b_i^t$ connecting the two marked points on boundary
component $i$ of $C_{0,3}^t$ with positive orientation\footnote{The orientation is indicated in Figure \ref{pantsdef}.
The starting point is determined either by the base point or by the end-point of the previous arc.}. 
\end{itemize}

We will assume that the point $y_0$
is located on the boundary circle $\de_n$ of $C_{0,n}$.
It will be useful to introduce the notation
$[ji^\nu]_t$ for the composite arcs $(b_i^t)^\nu \circ[ji]_t $,
$\nu\in\BZ$.

\subsubsection{The algorithm}\label{algorithm}

Using the results from Subsection \ref{fusbraid} and the definitions from
\ref{param} we may now formulate a simple algorithm for
calculating the result of the analytic continuation
$\CF(\,\al;\be\,|\,z\,|\,y_0\,|\,y\,)$ along $\chi_k^{}$.
We will use the geometrical set-up introduced in
Subsection \ref{param}, in particular the decomposition of the
paths $\chi_k^{}$ into a collection of arcs. Note that the basic building
blocks are close relatives of the moves introduced in \rf{moves}
such as
\begin{align} \label{fus'}
\psi_{s_2}(y)V_{\al_3,\al_1}^{\al_2}[v_2](z)&\,=\,
\psi_{s_2}(y)V_{\al_3+s_2\frac{b}{2},\al_1}^{\al_2}[v_2](z)\cdot\SV_{\al_3}^{\frac{1}{2}s_2}\notag\\
& \,=\,\sum_{s_1=\pm 1} F_{-s_2,s_1}^{[23]}\,
V_{\al_{3},\al_1}^{\al_2-s_1\frac{b}{2}}
\big[\psi_{s_1}(y-z) v_2\big](z)
\cdot\SV_{\al_3}^{\frac{1}{2}s_2}\,.
\end{align}
In this way
we find that the arcs $[ji^\nu]_t$
are represented by the matrices
\begin{subequations}
\begin{equation}
\SS_{[ji]}^{t}:=\SF_{[ji]}^{t}\cdot \ST^t_i\,,
\qquad
\SC_{[ji]}^{t,\nu}:=\SS_{[ji]}^{t}\cdot(\SB^t_i)^{\nu}\,,
\end{equation}
where
\begin{itemize}
\item $\SF_{[ji]}^{t}$ is
obtained from $F^{[ji]}$ by
replacing $\al_i\ra\al_i^t$, $i=1,2,3$ and
transposition\footnote{We are here
representing fusion and braid moves
by matrix multiplication from the right to be consistent with
\rf{monoact}. This differs from the conventions used in \cite{DGOT}
where multiplication from the left was used.
The matrices written below are therefore related to those of \cite{DGOT}
by transposition.},
\item $\ST^t_i$ is defined as
\begin{equation}
(\ST^t_i)_{s_1s_2}^{}\,=\,\de_{s_1,-s_2}^{}\,(\SV_t)^{\frac{1}{2}s_2}\,,
\end{equation}
where $\SV_t$ is the shift operator which shifts the variable $\alpha_t\equiv
\alpha_i^t$ as
defined in equation \rf{shiftdef}.
The operators $\SV_t$ act to the left in the product of matrices.
\item $\SB_{i}^t$ is the matrix with elements
\begin{equation}
(\SB_{i}^t)_{s_1s_2}^{}\,=\,\de_{s_1s_2}^{}\,
B_{s_1}^{}(\al_i)\,.
\end{equation}
\end{itemize}
\end{subequations}
Arcs $b_{i}^t$ will be
represented by the matrix $\SB_{i}^t$.
If $\chi_k^{}$ is a simple closed curve on $C_{0,n}$ starting and
ending at $y_0$ represented by the ordered
concatenation $\eta_1\circ \eta_2
\circ\dots\circ \eta_K$ of the
arcs defined above,
we will define \begin{equation}
\SM_{k}=
\SN_K\cdot \SN_{K-1}\cdot\dots\cdot \SN_1\,,
\end{equation}
where $\SN_k$ are the $2\times 2$-matrices associated to the arcs $\eta_k$.
We may thereby define the sought-for
collection of matrices
$\SM_k$, $k=1,\dots,n$ describing the action
of monodromies of the degenerate fields on spaces of
conformal blocks.

One should not forget that the resulting monodromy matrix is
operator-valued: it is a matrix which has elements
containing the operators $\SV_t$ shifting the parameters
$\be$.

\section{Solving the Riemann-Hilbert problem}\label{sec_solRHP}

We shall now specialize to $c=1$. For that case we shall
replace the parameters $\al_k$ and $\be_r$ by variables $m_k$ and $p_r$ giving
the conformal dimensions as $\De_{m_k}=m_k^2$ and $\De_{p_r}=p_r^2$,
for $k=1,\dots,n$ and $r=1,\dots,n-3$, respectively.

\subsection{The construction}\label{constr}

Let us now consider,
\begin{align}\label{cfbldef}
&\CF_{s's}\big(\,m;p\,|\,z\,|\,y_0\,|\,y\,\big):=\,
\big\langle\,m_{n}\,|\,\psi_{-s'}(y_0)\psi_{s}(y)\,|\,\Theta_{{s-s'}}\,\big\rangle\,,\\
& |\,\Theta_{\epsilon}\,\rangle\,=\,
V^{m_{n-1}}_{m_{n}+\frac{\epsilon}{2}\,,\,p_{n-3}}(z_{n-1})\dots V^{m_3}_{p_2,p_1}(z_3)\,V^{m_2}_{p_1,m_1}(z_2)\,|\,m_1\,\big\rangle\,,
\notag\end{align}
where $V^{m}_{p_2,p_1}(z)$ maps $\CV_{p_1}$ to $\CV_{p_2}$ and
$\psi_{s}(y)$ maps $\CV_{p}$ to $\CV_{p-s/2}$ for all $p$. We will from now
on assume that the vertex operators
$V^{m}_{p_2,p_1}(z)$ are normalized by \rf{normcond}
with $N(p_3,p_2,p_1)$ being chosen as
\begin{align}\label{Ndef}
&N(p_3,p_2,p_1)\,=\,\\
&=\,\frac{G(1+p_3-p_2-p_1)G(1+p_1-p_3-p_2)G(1+p_2-p_1-p_3)G(1+p_3+p_2+p_1)}{G(1+2p_3)G(1-2p_2)G(1-2p_1)}\,,
\notag\end{align}
where $G(p)$ is the Barnes $G$-function that satisfies $G(p+1)=\Ga(p)G(p)$.

Consider the matrix $\Psi(y;y_0)$ which has elements
\begin{subequations}\label{dualblocks}
\begin{equation}
\Psi_{s's}(y;y_0):=\,\frac{\pi s'(y_0-y)^{\frac12}}{\sin2\pi m_n}\frac{\big\langle\,m_{n}\,|\,\psi_{-s'}(y_0)\psi_{s}(y)\,|\,\Theta^{\rm D}_{{s-s'}}\,\big\rangle}{\big\langle\,m_{n}\,
|\,\Theta^{\rm D}_0\,\big\rangle}\,,
\end{equation}
where
\begin{equation}\label{FDdef}
|\,\Theta^{\rm D}_{\epsilon}(\si,\tau)\,\rangle:=
\sum_{\vec{n}\in\BZ^{N}}\,\prod_{r=1}^{N}\,e^{i{n_r}\tau_r}\,
|\,\Theta_{\epsilon}(\si+\vec{n})\,\rangle\,.
\end{equation}
\end{subequations}
We have introduced $N:=n-3$, and the summation is over
vectors $\vec{n}=(n_1,\dots,n_N)$ in $\BZ^N$. We claim that
$\Psi_{s's}(y;y_0)$ represents
the sought-for solution to the Riemann-Hilbert problem.
The proof of this statement is given in the following
subsections. At this point we only remark that the prefactor in (\ref{dualblocks}a)
ensures the normalization $\Psi(y_0;y_0)=1$.

The  observations above provide the input needed to
apply the reasoning presented in \cite{GIL} to show that the
isomonodromic tau-function is nothing but
\begin{equation}\label{taufourier}
\tau(\,z\,)\,=\,
\big\langle\,m_{n}\,|\,\Theta^{\rm D}_0\,\big\rangle\,.
\end{equation}
Our results for the case $n=4$ yield in particular a proof of the relation between
the tau function for Painlev\'e VI and Virasoro conformal blocks
discovered in \cite{GIL}.

\subsection{Existence of classical monodromies}\label{proof}


We may calculate the monodromies by the algorithm
formulated in Subsection \ref{algorithm} with input data
$F^{[ji]}_{s_1,s_2}$ and $B_{s}(\al)$ now given by
\begin{subequations}
\begin{align}\label{q-Gmatrix}
&F^{[ji]}_{s_1,s_2}=s_1\frac{\cos\pi(p_k+s_2p_j-s_1p_i)}{\sin2\pi p_j}
\,,\\
&B_{s}^{}(p)\,=\,e^{-\pi {\mathrm i} \,s\,p}\,.
\end{align}\end{subequations}
The operator
$\SV_t$ may now be represented as $\SV_t=e^{{\mathrm i}\sq_t}$,
where $\sq_t=i\frac{\pa}{\pa p_t}$.
Let us denote the resulting operator-valued monodromy matrices
by $\SM_{\ga}$.

We may now make a key observation: the monodromy matrices
$\SM_{\ga}$ have matrix elements that are rational functions
of $\SU_t=e^{2\pi i p_t}$ and $\SV_t$
which generate a {\it commutative} subalgebra of the
algebra of all operators\footnote{We mean operators acting on the
conformal blocks built from
$p_t$, $\pa_{p_t}$. }
acting on the space of conformal blocks.

In order to see that $\SM_{\ga}$ depends only on $\SV_t$
rather than $(\SV_t)^{\frac{1}{2}}$ let us note that
each curve of the cut system traversed on the way must be
crossed a second time before one can return to the starting point.
In a similar way one may see that $\SM_{\ga}$ depends on $p_t$
only via $U_t=e^{2\pi i p_t}$: the elements of the
matrices $F^{[ji]}_{s_1,s_2}$ are linear combinations
of the form $A e^{\pi i p_t}+B e^{-\pi ip_t}$.
As the product of matrices representing $\SM_{\ga}$
will always contain an even number of
matrices depending on a given variable $p_t$, it follows that
$\SM_{\ga}$ depends on $p_t$ only via $e^{2\pi ip_t}$.

But this means that the algebra generated by the matrix
elements of $\SM_{\ga}$ becomes classical (commutative) in the
limit $c\ra 1$!
This allows us to diagonalize the operator $\SV_t$ by taking
linear combinations
of the form \rf{FDdef}.
The generalized Fourier-transformation \rf{FDdef} diagonalizes $\SV_t$ with eigenvalue $e^{i\tau_t}$,
while $e^{2\pi ip_t}$ will act on $\Psi_{s's}$ by multiplication.
The matrix obtained from $\SM_{\ga}$ by means
of the transformation \rf{FDdef} will be denoted $M_{\ga}$.

\subsection{Calculation of monodromies}\label{Mcalc}

In order to formulate the rules for the calculation of the
monodromy matrices $M_{k}$,
let us assume without loss of generality
that the path connecting boundary component
$\de_n$ to $\de_k$ passes through the trinions $t_1, t_2,\ldots ,t_L$
in the given order, each trinion being traversed exactly once.
We claim that we may then calculate the monodromy matrices
$M_k$ as
\begin{equation}\label{M-calc}
M_{k}=\si_3\cdot
\big[C^{t_L,\nu_L}_{[j_Li_L]}\dots C^{t_1,\nu_1}_{[j_1i_1]}\big]^{-1}
\cdot
(\SB^{t_L}_k)^2
\cdot
\big[C^{t_L,\nu_L}_{[j_Li_L]}\dots C^{t_1,\nu_1}_{[j_1i_1]}\big]\cdot
\si_3,
\end{equation}
where $\si_3=\big(\begin{smallmatrix}
1 & \;\;0 \\ 0 & -1 \end{smallmatrix}\big)$ and
$C_{[ji]}^{t,\nu}$ is defined
as
\begin{subequations}\label{cnu}
\begin{equation}
C_{[ji]}^{t,\nu}:=\,F_{[ji]}^{t}\cdot (T\!B)^{t,\nu}_i\,,
\end{equation}
with matrices $F_{[ji]}^{t}$, and $(T\!B)_i^{t,\nu}$ defined
as
\begin{align}\label{cl-Fmatrix}
&(F_{[ji]}^t)_{s_1,s_2}=
s_2\frac{\sin\pi(p_k^t+s_2p_j^t-s_1p_i^t)}{\sin2\pi p_j^t}\,,\\
& \label{Trdef}
[(T\!B)_i^{t,\nu}]_{s_1s_2}^{}\,=\,\de_{s_1,-s_2}^{}\,{\mathrm i}^{s_2\nu}
e^{s_2\frac{\mathrm i}{2}\tau^t_i}e^{-\pi {\mathrm i}\,\nu \,s_2\,p^t_i}\,.
\end{align}
\end{subequations}

In order to derive these rules let us note that application of the
algorithm formulated in Subsection \ref{algorithm} will produce
monodromy matrices $\SM_k$ of the following form:
\begin{equation}\label{SM-calc}
\SM_{k}\,=\,
\big[\SC^{t_L,\nu_L}_{[j_Li_L]}\cdot\dots\SC^{t_1,\nu_1}_{[j_1i_1]}\big]^{-1}
\cdot
(\SB^{t_L}_k)^2
\cdot
\big[\SC^{t_L,\nu_L}_{[j_Li_L]}\cdot\dots\SC^{t_1,\nu_1}_{[j_1i_1]}\big]\,.
\end{equation}
The matrices
$\SC^{t,\nu}_{[ji]}$, $\nu\in\BZ$,
represent the contribution of the segments
connecting boundary component $i$ to $j$
in trinion $t$.
Recall that
$\SC^{t,\nu}_{[ji]}=\SS_{[ji]}^t\cdot (\SB_i^t)^\nu$.
The matrices $\SS_{[ji]}^t$ and $(\SS^{t}_{[ji]})^{-1}$
are explicitly given as
\begin{align*}
\SS^{t}_{[ji]}  &=\frac{1}{\sin 2\pi p^t_j}
\left(
\begin{matrix} -\sin\pi (p^t_k+p^t_j-p^t_i-\fr{1}{2}) &
\sin\pi (p^t_k-p^t_j-p^t_i+\fr{1}{2}) \\
-\sin\pi (p^t_k+p^t_j+p^t_i-\fr{3}{2}) &
\sin\pi (p^t_k-p^t_j+p^t_i-\fr{1}{2})
\end{matrix}\right)\left(
\begin{matrix} 0 & e^{-\frac{\mathrm i}{2}\sq^t_i} \\ e^{\frac{\mathrm i}{2}\sq^t_i} & 0
\end{matrix}\right)\notag\\
&=\frac{1}{\sin 2\pi p^t_j}\left(
\begin{matrix} -e^{\frac{\mathrm i}{2}\sq^t_i}\sin\pi (p^t_k-p^t_j-p^t_i) &
e^{-\frac{\mathrm i}{2}\sq^t_i}\sin\pi (p^t_k+p^t_j-p^t_i) \\
-e^{\frac{\mathrm i}{2}\sq^t_i}\sin\pi (p^t_k-p^t_j+p^t_i) &
e^{-\frac{\mathrm i}{2}\sq^t_i}\sin\pi (p^t_k+p^t_j+p^t_i)
\end{matrix}
\right),\\
(\SS^{t}_{[ji]})^{-1} &=\frac{1}{\sin 2\pi p^t_i}\left(
\begin{matrix} \sin\pi (p^t_k+p^t_j+p^t_i) e^{-\frac{\mathrm i}{2}\sq^t_i}
&
-\sin\pi (p^t_k+p^t_j-p^t_i)e^{-\frac{\mathrm i}{2}\sq^t_i}
\\
\sin\pi (p^t_k-p^t_j+p^t_i)e^{+\frac{\mathrm i}{2}\sq^t_i}
&
-\sin\pi (p^t_k-p^t_j-p^t_i)e^{+\frac{\mathrm i}{2}\sq^t_i}
\end{matrix}
\right).
\end{align*}
In order to calculate the effect of the
transformation \rf{FDdef} it is convenient to
move the operators $e^{\pm {\mathrm i}\sq_t}$ to the left  in \rf{SM-calc}.
To this aim let
us analyze the dependence of $\SM_k$ on $p_{i_a}\equiv p_{i_a}^{t_a}$
and the shift operator
$e^{{\mathrm i}\sq_a}$, where $\sq_a\equiv\sq_{i_a}^{t_a}$.
The dependence on $e^{{\mathrm i}\sq_a}$
can be made explicit by writing $\SM_k$ as
\begin{equation}\label{ordering}
\SM_k = \big[ \SC^{t_{a-2}\dots t_{1}}\big]^{-1}\!\cdot
\big[\SC^{t_{a-1},\nu_{a-1}}_{[j_{a-1}i_{a-1}]}\big]^{-1}\!\cdot
[\SB_{i_a}^{t_a}]^{-\nu_a}\!\cdot\SM_{k,a}'\cdot
(\SB_{i_a}^{t_a})^{\nu_a}\!\cdot
\SC^{t_{a-1},\nu_{a-1}}_{[j_{a-1}i_{a-1}]} \cdot
\big[ \SC^{t_{a-2}\dots t_{1}}\big]\,,
\end{equation}
where
\[
\SM_{k,a}':=\big[\SS^{t_a}_{[j_ai_a]}\big]^{-1}
\cdot\big[\SC^{t_L,\nu_L}_{[j_Li_L]}\cdots\SC^{t_{a+1},\nu_{a+1}}_{[j_{a+1}i_{a+1}]}\big]^{-1} \cdot
(\SB^{t_L}_k)^2 \cdot \big[\SC^{t_L,\nu_L}_{[j_Li_L]}\cdots\SC^{t_{a+1},\nu_{a+1}}_{[j_{a+1}i_{a+1}]}\big]\cdot\SS^{t_a}_{[j_ai_a]}\,.
\]
It is easy to see that the dependence of the
matrix $\SM_{k,a}'$ on $e^{{\mathrm i}\sq_a}$ is of the form
\[
\SM_{k,a}'\,=\,\left(\begin{matrix} \;\;\;\;\;\;\;\;(\sm_{k,a}')_{++}^{} & -e^{-{\mathrm i}\sq_a} \,
(\sm_{k,a}')_{+-}^{} \\ -e^{{\mathrm i}\sq_a} \,
(\sm_{k,a}')_{-+}^{} &
\;\;\;\;\;\;\;\;\;\;(\sm_{k,a}')_{--}^{} \end{matrix}
\right)\,,
\]
where $\sm_{k,a}'$ is the matrix one would obtain by replacing $\sq_a$
by $0$ and $\SF^{t_a}_{[j_ai_a]}$ by $F^{t_a}_{[j_ai_a]}$ from the very beginning. The extra minus sign is the result
of the application of the exchange relation
\[
\sin\pi (p_{k_a}+p_{j_a}+p_{i_a}) e^{-{\mathrm i}\sq_a}
=-e^{-{\mathrm i}\sq_a}\sin\pi (p_{k_a}+p_{j_a}+p_{i_a})\,.
\]

The only matrices  in \rf{ordering}
to the left of  $\SM_{k,a}'$ containing dependence on the variable
$p_{i_a}$ are
$[\SC^{t_{a-1},\nu_{a-1}}_{[j_{a-1}i_{a-1}]}]^{-1}$ and $[\SB_{i_a}^{t_a}]^{-\nu_a}$. The matrix elements
of both $\SS^{t_{a-1}}_{[j_{a-1}i_{a-1}]}$ and $\SB_{i_a}^{t_a}$ are both
anti-periodic under shifts $p_a\ra p_a+1$. Moving $e^{\pm{\mathrm i}\sq_a}$
through the product
$[\SC^{t_{a-1},\nu_{a-1}}_{[j_{a-1}i_{a-1}]}]^{-1}\cdot
[\SB_{i_a}^{t_a}]^{-\nu_a}$
will
for $a>1$ produce an extra sign $(-)^{1+\nu_a}$. This sign
is taken into account by means of the factor
${\mathrm i}^{s_2\nu_a}$ in \rf{Trdef}.
The extra sign $(-)^{1+\nu_a}$ should be replaced by $(-)^{\nu_a}$ in the case $a=1$.
This is taken into account by means
of conjugation with $\si_3$ in
\rf{M-calc}.

Calculating the trace functions $L_s^r$ and $L_{t}^r$ using the
algorithm above shows that
the parameters $(\si,\tau)$ coincide with those introduced in
Subsection \ref{Darboux}. The details are given in Appendix~\ref{appTF}.

\section{Non-Abelian fermionization}

It was shown in the work of Sato, Jimbo and Miwa that
the isomonodromic tau-functions can be represented in terms of
free fermion correlators. Our results give a ``bosonic'' representation
for the isomonodromic tau-functions in terms of Virasoro vertex
operators. In this section we will clarify the relation between
these two constructions
by showing that our construction is essentially
equivalent to a bosonic construction of twist fields creating
singularities with nontrivial monodromy. It seems natural
to regard our construction as the bosonization of the
fermionic construction of twist fields presented by
Sato, Jimbo and Miwa.

\subsection{Fermions from degenerate fields}

Let us introduce a free field $\vf_0$,
\[
\vf_0(w)\vf_0(z)\,\sim\,-\frac{1}{2}\log(w-z)\,.
\]
Note furthermore that we have
\begin{equation}
\De_{-b/2}\big|_{b=i}\,=\,\frac{1}{4}\,,\qquad
\De_{-b}\big|_{b=i}\,=\,1\,.
\end{equation}
Construct the fields
\begin{equation}
\Psi_s(z):=\,e^{i\vf_0(z)}\psi_s(z)\,,\qquad
\bar{\Psi}_s(z):=e^{-i\vf_0(z)}\psi_{-s}(z)\,.
\end{equation}
These fields have the OPE
\begin{subequations}
\label{FF-OPE}
\begin{align}
\Psi_s(w)\Psi_{s'}(z)\,& \sim\,{\rm regular}\,,\\
\Psi_s(w)\bar\Psi_{s'}(z)\,& \sim\,\frac{\de_{s,s'}}{w-z}\,.
\end{align}
\end{subequations}
This means that the fields $\Psi_s(w)$, $\bar\Psi_s(w)$ generate
a representation of the fermionic vertex operator algebra ${\mathfrak F}$.
The action of these fields can be restricted to the spaces
$\CF_{\si,\tau}$, defined as
\begin{equation}
\CF_{\si,\tau}:
=\,\bigoplus_{\substack{k,l\in\frac{1}{2}\BZ \\k+l\in\BZ}}
\CF_{\si,\tau}^{[k,l]}\,,\qquad \CF_{\si,\tau}^{[k,l]}:=
\CV_{\si-{k}}^{}\ot \CF^{}_{\tau+{l}}\,.
\end{equation}
with $\CF^{}_{\tau}$ being the free boson Fock space
with eigenvalue $\tau$ for the zero mode of $\pa\vf_0$.
Note that the action of $\Psi_s(z)$, $\bar{\Psi}_s(z)$
shifts $k+l$ by an integer amount.
In order to get a label for inequivalent representations of \rf{FF-OPE}
we may restrict $\si$ and $\tau$ to
$0\leq\Re(\si)<1/2$ and $0\leq\Re(\tau)<1$, respectively.

The restriction of $\Psi_s(z)$, $\bar{\Psi}_s(z)$ to
$\CF_{\si,\tau}$ has monodromy
\begin{equation}
\Psi_s(e^{2\pi i}z)\,=\,e^{2\pi i(\tau-s\si)}\,\Psi_s(z)\,,\qquad
\bar\Psi_s(e^{2\pi i}z)\,=\,e^{-2\pi i(\tau-s\si)}\,\bar\Psi_s(z)\,.
\end{equation}
Other representations of the fermionic vertex operator
algebra \rf{FF-OPE} can be defined by taking linear combinations
\begin{equation}\label{chofbasis}
\Phi_s(z):=\,\sum_{t=\pm}C_{st}\Psi_t(z)\,,\qquad
\bar\Phi_s(z):=\,\sum_{t=\pm}(C^{-1})_{st}\bar\Psi_t(z)\,,
\end{equation}
for any element $C$ of ${\rm GL}(2)$. The representation is characterized
by the ${\rm GL}(2)$-monodromy
\begin{equation}\label{GL2-mono}
\begin{aligned}
&\Phi_s(e^{2\pi i}z)\,=\,\sum_{t=\pm}M_{st}\Phi_t(z)\,,\qquad
\bar\Phi_s(e^{2\pi i}z):=\,\sum_{t=\pm}(M^{-1})_{st}\bar\Phi_t(z)\,,\\
&{\rm where}\;\;
M\,=\,C\cdot e^{2\pi i D}\cdot C^{-1}\,,\qquad
D:={\rm diag}(\tau-\si,\tau+\si)\,.
\end{aligned}\end{equation}
It seems natural to consider equivalence classes of representations defined
by identifying representations related by the similarity transformation
\rf{chofbasis}. Slightly abusing notations we will denote the
representations characterized by monodromy of the form
\rf{GL2-mono} by $\CF_{\si,\tau}$.

It will be useful to decompose $\CF_{\si,\tau}$ as
\begin{equation}
\CF_{\si,\tau}\,=\,\CF_{\si,\tau}^{0}\oplus \CF_{\si,\tau}^{1/2}\,,
\quad{\rm where}\quad
\CF_{\si,\tau}^{\ep}:=\,
\bigoplus_{\substack{k\in\BZ,\;l\in\frac{1}{2}\BZ\\k+l\in\BZ}}
\CF_{\si,\tau}^{[k+\ep,l]}\,,
\end{equation}
assuming that $\ep\in\frac{1}{2}\BZ_2$.
The action of a field $\Psi_s(w)$, $\bar\Psi_s(w)$ maps
$\CF_{\si,\tau}^{0}$ to $\CF_{\si,\tau}^{1/2}$ and vice-versa.

\subsection{Chiral vertex operators for free fermion representations}

Let us then define the vertex operators
\begin{equation}
\Phi_{\si_3,\tau_3\;;\;\si_1,\tau_1}^{\si_2,\tau_2\;;\;\ep_2,\,q_3}(z):
\CF_{\si_1,\tau_1}\ra
\CF_{\si_3,\tau_3}\,,
\end{equation}
by defining their
action on arbitrary vectors $v_1\in \CF_{\si_1,\tau_1}^{[k_1+\ep_1,l_1]}$
to be
\begin{equation}
\Phi_{\si_3,\tau_3\;;\;\si_1,\tau_1}^{\si_2,\tau_2\;;\;\ep_2,\,q_3}(z)\,v_1:=
e^{2i\tau_2^{}\vf_0^{}(z)}
\sum_{n\in\BZ} \,e^{inq_3}\,
V_{\si_3-[\ep_1+\ep_2]+n\;;\;\si_1-k_1-\ep_1}^{\si_2-\ep_2}(z)\,v_1
\,;
\end{equation}
we assume that $\tau_3=\tau_2+\tau_1$, and define
$[\ep]=0$ if $\ep\in\BZ$, $[\ep]=1/2$ if $\ep\in\BZ+\frac{1}{2}$.
The definition is such that the restriction of
$\Phi_{\si_3,\tau_3\;;\;\si_1,\tau_1}^{\si_2,\tau_2\;;\;\ep_2,\,q_3}(z)$ to
the subspace $\CF_{\si_1,\tau_1}^{\ep_1}$ of $\CF_{\si_1,\tau_1}$
yields an operator with image
contained in the subspace $\CF_{\si_3,\tau_3}^{\ep_1+\ep_2}$ of
$\CF_{\si_3,\tau_3}$. This selection rule expresses conservation
of the quantum number $\ep\in\frac{1}{2}\BZ_2$.

The relations \rf{braid} combined with the standard braid relations of normal ordered exponentials imply the following exchange relations between the vertex
operators $\Phi_{\si_3,\tau_3\;;\;\si_1,\tau_1}^{\si_2,\tau_2\;;\;\ep_2,\,q_3}(z)$
and the fermion fields $\Psi_s(w)$,
\begin{equation}\label{exchn}
\begin{aligned}
\Psi_s(w\pm i0)\,
\Phi_{\si_3,\tau_3\;;\;\si_1,\tau_1}^{\si_2,\tau_2\;;\;\ep_2,q_3}(z)\,
=\,\Phi_{\si_3,\tau_3\;;\;\si_1,\tau_1}^{\si_2,\tau_2\;;\;\ep_2,q_3}(z)\,
\sum_{t=\pm}
\Psi_{t}(w\pm i0)\,B^\pm(q_3)_{t,s}\,,\\
\bar\Psi_s(w\pm i0)\,
\Phi_{\si_3,\tau_3\;;\;\si_1,\tau_1}^{\si_2,\tau_2\;;\;\ep_2,q_3}(z)\,
=\,\Phi_{\si_3,\tau_3\;;\;\si_1,\tau_1}^{\si_2,\tau_2\;;\;\ep_2,q_3}(z)\,
\sum_{t=\pm}
\bar\Psi_{t}(w\pm i0)\,\bar{B}^\pm(q_3)_{t,s}\,;
\end{aligned}\end{equation}
The matrices $B_{t,s}^\pm(q_3)$ and
$\bar{B}_{t,s}^\pm(q_3)$ are explicitly given as
\begin{equation}\begin{aligned}
&B^\pm(q_3)_{t,s}\,=\,e^{\pm\pi i \tau_2}\,
e^{\mp\pi i(s\si_3-t\si_1)}\,
e^{i([\ep_3+\frac{1}{2}]-\ep_3-\frac{s}{2})q_3}\,
F_{st}^{[13]}(\si_3,\si_2-\ep_2,\si_1)\,,\\
& \bar{B}^\pm(q_3)_{t,s}\,=\,e^{\mp\pi i \tau_2}\,
e^{\pm\pi i(s\si_3-t\si_1)}\,
e^{i([\ep_3+\frac{1}{2}]-\ep_3+\frac{s}{2})q_3}\,
F_{-s,-t}^{[13]}(\si_3,\si_2-\ep_2,\si_1)\,.
\end{aligned}
\end{equation}
The exchange relations \rf{exchn}
express the fact that
$\Phi_{\si_3,\tau_3\;;\;\si_1,\tau_1}^{\si_2,\tau_2\;;\;\ep_2,q_3}(z)$
is an intertwiner between the
representations $\CF_{\si_1,\tau_1}$ and $\CF_{\si_3,\tau_3}$ of the
free fermion algebra $\mathfrak F$. It also follows from these
observations that the vertex operators
$\Phi_{\si_3,\tau_3\;;\;\si_1,\tau_1}^{\si_2,\tau_2\;;\;\ep_2,q_3}(z)$
represent twist fields:
They create states in which the
fermions $\Psi_s(z)$ have
monodromy $B^-(q_3)(B^+(q_3))^{-1}$ around $z$.

An important consequence of \rf{exchn} is the fact that matrix elements of
compositions of the vertex operators
$\Phi_{\si_3,\tau_3\;;\;\si_1,\tau_1}^{\si_2,\tau_2\;;\;\ep_2,\,q_3}(z)$
such as
\begin{equation}
\langle\,e_{\si_4,\tau_4}^{\ep_4}\,|\,
\Phi_{\si_4,\tau_4\;;\;\si,\tau}^{\si_3,\tau_3\;;\;\ep_3,\,q_4}(z_3)\,
\Phi_{\si,\tau\;\;\;\; ;\; \si_1,\tau_1}^{\si_2,\tau_2\;;\;\ep_2,\,q_3}(z_2)
\,|\,e_{\si_1,\tau_1}^{\ep_1}\,\rangle
\end{equation}
represent conformal blocks for the free fermion algebra $\mathfrak F$.
$|\,e_{\si,\tau}^{\ep}\,\rangle$ is the product of highest weight vertors
in $\CV_{\si-\ep}^{}\ot \CF^{}_{\tau}$.
It follows from the conservation of the quantum number $\ep$ that
such conformal blocks
are non-vanishing only if $\ep_4=\ep_1+\ep_2+\ep_3\;{\rm mod}\;\,1$.
Conservation of the zero mode of $\vf_0$ implies furthermore that
$\tau_4=\tau_1+\tau_2+\tau_3$.

The free fermion conformal blocks
factorize as
\begin{align}\langle\,e_{\si_4,\tau_4}^{\ep_4}\,|\,
& \Phi_{\si_4,\tau_4\;;\;\si,\tau}^{\si_3,\tau_2\;;\;\ep_3,q_4}(z_3)\,
\Phi_{\si,\tau;\; \si_1,\tau_1}^{\si_2,\tau_2\;;\;\ep_2,q_3}(z_2)
\,|\,e_{\si_1,\tau_1}^{\ep_1}\,\rangle_{\rm FF}^{}\,=\,\\[1ex]
&\,=\,
\langle\,\tau_4\,|\,e^{2i\tau_3\vf_0(z_3)}\,e^{2i\tau_2\vf_0(z_2)}\,
|\,\tau_1\,\rangle^{}_{0}\notag\\
&\hspace{.75cm}\times
\sum_{n\in\BZ} e^{inq}\langle\,\si_4-\ep_4\,|\,
V_{\si_4-\ep_4,\;\si-[\ep_1+\ep_2]+n}^{\si_3-\ep_3}(z_3)\,
V_{\si-[\ep_1+\ep_2]+n;\;\si_1-\ep_1}^{\si_2-\ep_2}(z_2)
\,|\,\si_1-\ep_1\,\rangle_{\rm Liou}^{}\,.
\notag\end{align}
The factor in the last line was previously
identified as the tau-function associated to isomonodromic
deformations of ${\rm SL}(2)$-connections, the free-field conformal block in the second
line is nothing but the multiplier needed to get the tau-functions
associated to the ${\rm GL}(2)$-connections.

\section{Examples}
We now look at some of the applications of the above general formalism to
the theory of monodromy preserving deformations. We start by providing a CFT derivation
of the Jimbo's asymptotic formula \cite{Ji} for the tau function of Painlev\'e~VI equation. Next we show
how the known algebro-geometric solutions of the Schlesinger system on $C_{0,n}$ \cite{kk} arise
from conformal blocks of the Ashkin-Teller critical model \cite{Za,ZZ}.

\subsection{Painlev\'e VI and Jimbo's formula}
\begin{figure}[h]
\epsfxsize13.5cm
\centerline{\epsfbox{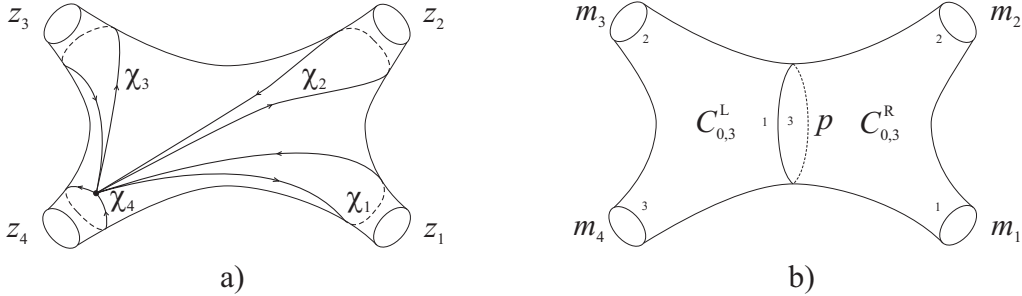}}
\caption{\it Basis of loops of $\pi_1(C_{0,4})$ and the decomposition $C_{0,4}=C_{0,3}^L\cup C_{0,3}^R$.}
\label{c04}\vspace{.3cm}
\end{figure}

Consider the simplest nontrivial case of four punctures. The fundamental
group $\pi_1(C_{0,4})$ is isomorphic to free group of rank~3. Let $\chi_1^{},\ldots,\chi_4^{}$ be
the four loops shown in Figure~\ref{c04}a, then
\beq
\pi_1(C_{0,4})=\left\langle \chi_1^{},\chi_2^{},\chi_3^{},\chi_4^{}\,|\,\chi_1^{}\circ\chi_2^{}\circ\chi_3^{}\circ\chi_4^{}=1\right\rangle.
\eeq
We denote by $M_1,\ldots ,M_4\in{\rm SL}(2,\BC)$ the monodromy matrices associated to these loops, satisfying
$M_4M_3M_2M_1=1$. Conjugacy classes of irreducible representations of $\pi_1(C_{0,4})$ in ${\rm SL}(2,\BC)$ are uniquely specified by
seven invariants
\begin{subequations}
\beqa
&L_k=\operatorname{Tr} M_k=2\cos2\pi m_k,\qquad k=1,\ldots,4,\\
&L_s=\operatorname{Tr} M_1 M_2,\qquad L_t=\operatorname{Tr} M_2 M_3,\qquad L_u=\operatorname{Tr} M_1 M_3,
\eeqa
\end{subequations}
generating the algebra of invariant polynomial functions on $\operatorname{Hom}\left(\pi_1(C_{0,4}),{\rm SL}(2,\BC)\right)$. These
traces satisfy the quartic equation
\begin{align}
 \label{JFR}
& L_1L_2L_3L_4+L_sL_tL_u+L_s^2+L_t^2+L_u^2+L_1^2+L_2^2+L_3^2+L_4^2=\\
 &\nonumber \quad=\left(L_1L_2+L_3L_4\right)L_s+\left(L_2L_3+L_1L_4\right)L_t+\left(L_1L_3+L_2L_4\right)L_u+4.
\end{align}

The affine algebraic variety defined by (\ref{JFR}) is the character variety of $C_{0,4}$. For every choice
of $m_1,\ldots,m_4$, it defines a cubic surface in $\BC^3$ in the variables $L_s$, $L_t$, $L_u$. If we further
fix the trace function $L_s=2\cos2 \pi \sigma$, the resulting quadric in $L_t$, $L_u$ admits rational parameterization~\cite{Ji}
\begin{subequations}
\label{Ji_par1}
\begin{align}
&\left(L_s^2-4\right)L_t=\,D_{t,+}\mathsf{s}\,+D_{t,-}\mathsf{s}^{-1}\,+D_{t,0},\\
&\left(L_s^2-4\right)L_u=D_{u,+}\mathsf{s}+D_{u,-}\mathsf{s}^{-1}+D_{u,0}.
\end{align}
\end{subequations}
with coefficients  given by
\begin{subequations}
\label{Ji_par2}
\begin{align}
&D_{t,0}=L_s\left(L_1L_3+L_2L_4\right)-2\left(L_1L_4+L_2L_3\right),\\
&D_{u,0}=L_s\left(L_2L_3+L_1L_4\right)-2\left(L_1L_3+L_2L_4\right),\\
&D_{t,\pm}=16\prod_{\epsilon=\pm}\sin\pi\left(m_2\mp \sigma+\epsilon m_1\right)\sin\pi\left(m_3\mp \sigma+\epsilon m_4\right),\\
&D_{u,\pm}=-D_{t,\pm}e^{\mp 2\pi i \sigma}.
\end{align}
\end{subequations}

The local coordinates $(\sigma,\mathsf{s})$ parameterize the space of ${\rm SL}(2,\BC)$-representations of $\pi_1(C_{0,4})$
with fixed local monodromy exponents $m_1,\ldots,m_4$.
Let us connect this pair to the parameters used in the conformal block representation of the fundamental matrix $Y(y)$.

The Riemann surface  $C_{0,4}$
is glued from two three-holed spheres $C_{0,3}^L$, $C_{0,3}^R$  as shown in Figure~\ref{c04}b.
The local coordinates $(p,\tau)$ associated to this pants decomposition parameterize trace functions via
(\ref{classWT})--(\ref{Cepdef})
(as well as their counterparts for $L_u$).
Comparing these expressions with (\ref{Ji_par1})--(\ref{Ji_par2}), we find
that
\beq\label{Jps}
\sigma=p,\qquad\mathsf{s}=
\frac{\sin\pi\left(\sigma-m_1+m_2\right)\sin\pi\left(\sigma+m_3-m_4\right)}{
\sin\pi\left(\sigma-m_1-m_2\right)\sin\pi\left(\sigma-m_3-m_4\right)}\,e^{i\tau}.
\eeq

Going back to the Schlesinger equations (\ref{Schlesinger}), note that
three regular singularities $z_1$, $z_3$, $z_4$ can be brought to $0$, $1$ and $\infty$ using
M\"obius transformations. The Schlesinger system then reduces to Painlev\'e VI equation
\begin{align}
&-\frac12\left(z(z-1)\zeta''\right)^2=\\
&\nonumber=\operatorname{det}
\left(\begin{array}{ccc}
2m_1^2 & z\zeta'-\zeta & \zeta'+m_1^2+m_2^2+m_3^2-m_4^2 \\
z\zeta'-\zeta & 2m_2^2 & (z-1)\zeta'-\zeta \\
\zeta'+m_1^2+m_2^2+m_3^2-m_4^2 & (z-1)\zeta'-\zeta & 2m_3^2
\end{array}\right),
\end{align}
satisfied by the logarithmic derivative of the tau function
\beq
\zeta(z)=\displaystyle z (z-1)\frac{d}{dz}\ln \tau.
\eeq
Here $\displaystyle z=\frac{(z_2-z_1)(z_4-z_3)}{(z_3-z_1)(z_4-z_2)}$ denotes the cross-ratio of the singular points.

In the case of $C_{0,4}$, the representation (\ref{taufourier})
of $\tau(z)$ as a Fourier transform of the $c=1$ Virasoro
conformal block  is more explicitly written as
\beq
\label{tauc04}
\tau(z)=\sum_{n\in\BZ}\langle\, m_4\,|\,V_{m_4,p+n}^{m_3}\left(1\right)V_{p+n,m_1}^{m_2}\left(z\right)|\,m_1\rangle\,e^{in\tau}.
\eeq
Assuming without loss of generality that $-\frac12<\Re p<\frac12$, letting $z\rightarrow 0$ in the last formula,
and taking into account the normalization (\ref{Ndef}) of the chiral vertex operators,
we deduce the asymptotics
\begin{align}
\nonumber\tau(z)=&\sum_{n=0,\pm1}N(m_4,m_3,p+n)N(p+n,m_2,m_1)e^{in\tau}z^{(p+n)^2-m_1^2-m_2^2}+\\
&+O\left(z^{p^2-m_1^2-m_2^2+1}\right).
\end{align}
This is equivalent to the famous Jimbo's asymptotic formula \cite[Theorem~1.1]{Ji} expressing the critical
behavior of the Painlev\'e VI tau function in terms of monodromy data. The relation of Jimbo parameters to ours is
given by (\ref{Jps}).

\subsection{Algebro-geometric solutions of the Schlesinger system}
\begin{figure}
\epsfxsize10.5cm
\centerline{\epsfbox{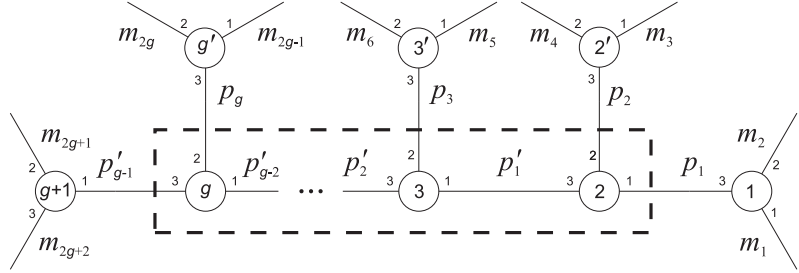}}
\caption{\it Labeling of pairs of pants for the conformal block $\mathcal{B}\left({m}\,|\,{p},{p}'\,|\,{z}\right)$.}
\label{AGblock}\vspace{.3cm}
\end{figure}

Consider the pants decomposition of  $C_{0,2g+2}$ schematically depicted in Figure~\ref{AGblock}, and denote by
$\mathcal{B}\left({m}\,|\,{p},{p}\,'\,|\,{z}\right)$
the corresponding $c=1$ conformal block. Its external legs are combined into $g+1$ pairs. The momenta obtained by
fusing different pairs are connected to a ``black box''. Its internal structure is not essential for the final result.
However, to fix the notations, we will choose it in a particular way and parameterize it by $g-2$ internal momenta $p_1',\ldots,p_{g-2}'$.

As explained in Section~\ref{sec_solRHP}, summation of conformal blocks over integer shifts of momenta
gives an isomonodromic tau function of the Schlesinger system,
\beq\label{AGsum1}
\tau(z)=\sum_{{n}\in\BZ^{g}}\sum_{{n}'\in \BZ^{g-1}}
\mathcal{B}\left({m}\,|\,{p}+{n},{p}'+{n}'\,|\,{z}\right)
e^{i{n}\cdot {\tau}+i{n}'\cdot {\tau}'}.
\eeq
 The variables ${p}$, ${p}'$,
${\tau}$, ${\tau}'$ provide a set of local coordinates on the $(4g-2)$-dimensional space of monodromy data.

Let us impose a free-field-like conservation constraint on momenta of the unshifted conformal block
 at each vertex inside the box. These conditions determine the black box momenta ${p}'={p}'[{p}]$ in terms
 of ${p}$. Explicitly,
 \beq
 \nonumber
 p_{k}'[{p}]=p_{k-1}'[{p}]+p_{k+1},\qquad p_0'[{p}]\equiv p_1.
 \eeq
 Also, for $k=1,\ldots,g-1$ we define
 \beq
 \nonumber\ell_k=n'_{k}-n'_{k-1}-n_{k+1},\qquad n'_0\equiv n_1.
 \eeq

 Since Barnes $G$-function
 vanishes at non-positive integer values of the argument, the form  of the
 normalization coefficient (\ref{Ndef}) restricts the sum (\ref{AGsum1}) to the domain  $\ell_1,\ldots,\ell_{g-1}\geq 0$.
 In the limit
 \begin{subequations}\label{taulimit}
 \beqa
 &\tau_j\rightarrow -i\infty,\qquad\tau'_{k}\rightarrow i \infty,\\
 &\tau_j+\sum_{k=j}^{g}\tau'_{k-1}\rightarrow \xi_j,\qquad
  \tau'_0\equiv 0,\qquad j=1,\ldots, g,
 \eeqa
 \end{subequations}
  this sum further reduces to the values $\ell_1=\ldots=\ell_{g-1}=0$. We thus get a $2g$-parameter family
  of tau functions
 \beq\label{AGsum2}
\tau(z)=\sum_{{n}\in\BZ^{g}}
\mathcal{B}\left({m}\,|\,{p}+{n},{p}'[{p}+{n}]\,|\,{z}\right)
e^{ i{n}\cdot {\xi}}.
\eeq
Notice that at each of $g-1$ internal vertices of conformal blocks which appear in (\ref{AGsum2}), the corresponding momenta satisfy the same conservation conditions as in the unshifted case.

\begin{figure}
\epsfxsize10.5cm
\centerline{\epsfbox{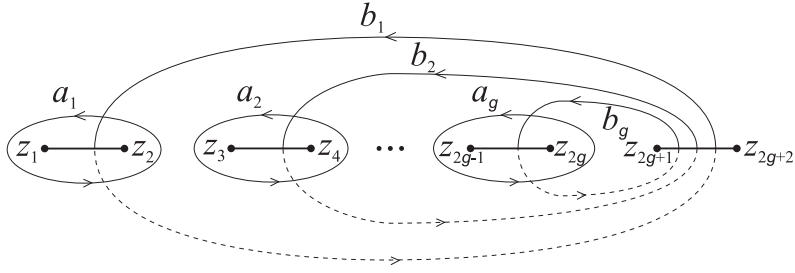}}
\caption{\it Canonical homology basis on $\Sigma$.}
\label{abcycles}\vspace{.3cm}
\end{figure}

Conformal blocks of this form with ${m}={m}_{\text{AT}}\equiv\left(\frac{1}{4},\ldots,\frac{1}{4}\right)$ describe correlation functions of the Ashkin-Teller critical model
\cite{Za,ZZ}. They can be expressed in terms of certain quantities associated
to the hyperelliptic curve $\Sigma$ of genus $g$ defined by
\beqa
\lambda^2=\prod_{k=1}^{2g+2}\left(y-z_k\right).
\eeqa
Let us fix the canonical homology basis of $a$- and $b$-cycles on $\Sigma$ as shown in Figure~\ref{abcycles}. The $g$-dimensional space of holomorphic $1$-forms on $\Sigma$ is spanned by
\beq
\nonumber d\omega_k=\frac{y^{k-1} dy}{\lambda},\qquad k=1,\ldots,g.
\eeq
The $g\times g$ matrices of $a$- and $b$-periods
\beq
{a}_{jk}=\oint_{a_k}d\omega_j,\qquad
{b}_{jk}=\oint_{b_k}d\omega_j,
\eeq
determine the symmetric period matrix $\Omega={a}^{-1}{b}$ of $\Sigma$. The hyperelliptic Riemann theta function
with characteristics $[\,{p}, {q} \,]\in\BC^{2g}$  is defined as the following series:
\beq
\label{thetadef}
\theta[\,{p}, {q} \,]\bigl({x}\, |\, \Omega\bigr)=
\sum_{{n}\in \BZ^g}e^{\pi i ({n}+{p})\cdot\Omega\cdot({n}+{p})
+2\pi i ({n}+{p})\cdot ({x}+{q})}.
\eeq
Even characteristics $[\,{p}_S, {q}_S \,]$ correspond to its non-trivial half-periods and are indexed by
partitions $S=\{\{z_{\alpha_1},\ldots,z_{\alpha_{g+1}}\},\{z_{\beta_1},\ldots,z_{\beta_{g+1}}\}\}$ of the set of ramification points into two subsets of equal size.

In this notation, the Ashkin-Teller conformal block is given by \cite{Za,ZZ}
\begin{align}
&\mathcal{B}\left({m}_{\text{AT}}\,|\,{p},{p}'\left({p}\right)|\,{z}\right)=
\mathcal{G}(p)\mathcal{K}(z)\,
\frac{e^{i\pi {p}\cdot\Omega\cdot{p}}}{\theta[\,{p}_S, {q}_S \,]\bigl({0}\, |\, \Omega\bigr)},\\
&\mathcal{G}(p)=\frac{\cos\pi p'_{g-1}}{\pi^{1-g/2}}\frac{{\hat{G}}^2(p'_{g-1}+\frac12)}{\prod_{k=1}^{g}{\hat{G}}^2(p_k)},\\
&\mathcal{K}(z)=\left(\frac{\prod_{j<k}^{g+1}(z_{\alpha_j}-z_{\alpha_k})\prod_{j<k}^{g+1}(z_{\beta_j}-z_{\beta_k})}{
\prod_{j,k}^{g+1}(z_{\alpha_j}-z_{\beta_k})}\right)^{\frac18}.
\end{align}
Here we denote $\hat{G}(p)=\frac{G(1+p)}{G(1-p)}$. The prefactor $\mathcal{G}(p)$ comes from our normalization
(\ref{Ndef}) of the chiral vertex operators. Taking into account the recurrence relation $\hat{G}\left(p+1\right)=-\pi\left(\sin\pi p\right)^{-1}\hat{G}\left(p\right)$, we see that
the sum (\ref{AGsum2}) reduces to the theta function series (\ref{thetadef}), so that
\beq
\tau(z)=\operatorname{const}\cdot\, \mathcal{K}(z)\,
\frac{\theta[\,{p}, {q} \,]\bigl({0}\, |\, \Omega\bigr)}{\theta[\,{p}_S, {q}_S \,]\bigl({0}\, |\, \Omega\bigr)},
\eeq
with $e^{2\pi i q_k}\equiv -\frac{\sin^2\pi p_k}{\cos^2 \pi p'_{g-1}}e^{i\xi_k}$.
We thus reproduce the $2g$-parameter family of tau functions found in \cite{kk}. The elliptic case $g=1$ corresponds
to Picard solutions of Painlev\'e VI.

At last let us compute the actual monodromy matrices for $m=m_{\text{AT}}$ applying the rules formulated in Subsection~\ref{Mcalc}.
Up to overall conjugation, one has
\begin{align}
M_{2k-1}&=\big[C_{[13]}^{k',1} C^{[k,\ldots, g]} C_{[13]}^{g+1,1}\big]^{-1}
\big(B^{k'}_1\big)^2 \,C_{[13]}^{k',1} C^{[k,\ldots, g]} C_{[13]}^{g+1,1},\\
M_{2k}\;\; &=\big[C_{[23]}^{k',0} C^{[k,\ldots, g]} C_{[13]}^{g+1,1}\big]^{-1}
\big(B^{k'}_2\big)^2 \,C_{[23]}^{k',0} C^{[k,\ldots, g]} C_{[13]}^{g+1,1},
\end{align}
with $C^{[k,\ldots , g]}=C_{[23]}^{k,-1}C_{[13]}^{k+1,0}\ldots C_{[13]}^{g,0}$ and $k=2,\ldots,g$. The conservation
of momenta at the vertices $k,\ldots,g-1$ implies that all matrices in the product $C^{[k,\ldots , g]}$ are lower triangular.
This enables one to explicitly calculate the monodromies in the limit (\ref{taulimit}). Again up to conjugation, the result
is
\begin{align}
M_k=\left(\begin{array}{cc} 0 & i\mu_k^{-1} \\ i\mu_k & 0\end{array}\right),\qquad k=1,\ldots,2g+2,
\end{align}
with $\mu_{2g+1}=e^{2\pi i p'_{g-1}}$, $\mu_{2g+2}=1$ and
\begin{align}
\mu_{2k-1}=e^{2\pi i (p'_{k-2}+ q_k)},\qquad
\mu_{2k}=-e^{2\pi i (p'_{k-1}+ q_k)}.
\end{align}
 Note in particular that in the chosen basis the products $M_{2k-1}M_{2k}$ and $M_{2k}M_{2k+1}$ are given by diagonal matrices,
 cf \cite[Theorem 3.2]{kk}.

\section{Outlook}

To conclude we will discuss some further applications and possible directions of future research
suggested by our results.

\subsection{Possible applications to the study of $\mathcal{N}=2$ supersymmetric gauge theories}

Our results appear to have interesting implications for the study of
a certain class of $4D$ $\mathcal{N}=2$ supersymmetric
gauge theories which is nowadays often called class $\CS$. The gauge theories $\CG_C$
in  class $\CS$ are associated to Riemann surfaces $C$, possibly with $n$ punctures.
The so-called instanton partition functions \cite{LNS,MNS1,MNS2,N,NO}
carry important non-perturbative information about the physics
of such gauge theories, including the
complete description of their low-energy physics via Seiberg-Witten
theory \cite{N,NO}.
Out of the instanton partition functions one may form
the so-called dual instanton partition functions by means of
a generalization of the Fourier series \cite{N,NO}.

It was observed in \cite{N,LMN,NO} that the
dual instanton partition functions of some
supersymmetric gauge theories from class $\CS$ have free fermion representations,
and therefore represent tau-functions for certain integrable
equations. Considerations of the geometric engineering of
such gauge theories within string theory have led to the
suggestion that the dual instanton partition functions of the 
gauge theories from
class $\CS$ should be related to the partition functions of chiral free fermion
theories on suitable Riemann surfaces \cite{N,ADKMV,DHSV,DHS}\footnote{The first proposal in this direction was formulated in \cite[Section 4.3]{N}.
The relations between the topological vertex
and free fermion theories discussed in \cite{ADKMV} 
imply general relations between topological string partition 
functions of toric Calabi-Yau manifolds, tau-functions and
theories of free fermions on certain Riemann surfaces; possible
implications for four-dimensional gauge theories were
discussed more explicitly 
in \cite{DHSV,DHS}. In some of the earlier references cited above, it was 
proposed that the relevant theory of
free fermions is defined on the Seiberg-Witten curve $\Sigma$ 
which for theories of class $\CS$ is a branched cover of the curve 
$C$ defining $\CG_C$.}. 
More recently it was  proposed in \cite{CNO} that the relevant
theory of chiral free fermions is defined on the Riemann surface
$C$ specifying the gauge theory $\CG_C$.
These relations were called BPS-CFT correspondence in \cite{CNO}.

In another important recent development it was found that the
instanton partition functions of these supersymmetric
gauge theories are related to the conformal blocks of the Toda
conformal field theories, in the simplest case the
Liouville theory \cite{AGT}. The correspondence between
instanton partition functions and Liouville conformal blocks
is called the AGT-correspondence.

However, up to now it was not clear how exactly BPS-CFT-corres\-pondence and
AGT-correspondence are related.
Our paper provides a basis for understanding these connections by
establishing a direct relation between the conformal field theory of chiral free fermions
on a Riemann sphere with $n$ punctures $C_{0,n}$ on the one hand, and
the conformal blocks of Liouville theory at $c=1$ on $C_{0,n}$ on the other hand.
Our result opens the interesting perspective to derive the $c=1$ case of the
AGT-correspondence from
the BPS-CFT-correspondence. It would suffice to characterise the relevant
$\bar{\partial}_E$-operators whose determinants should represent the dual instanton partition
function according the BPS-CFT-correspondence more precisely.  To this aim it may be
convenient to use the language proposed in \cite{DHS}. The connection between the
relevant determinants of $\bar{\partial}_E$-operators and the isomonodromic
tau-functions studied in this paper should then follow from the results of \cite{P}.
To complete the derivation of the AGT-correspondence for $c=1$ from the
BPS-correspondence it will suffice to observe that the Fourier-transformation appearing in
the
relation \rf{dualblocks} between conformal blocks and tau-functions is exactly the transformation from instanton partition functions to the dual instanton partition functions.

\subsection{Verlinde loop operators and quantisation of $\CM_{\rm flat}(C)$}

For $c\neq1 $ one may use the operator-valued monodromies constructed in Section \ref{q-mono} to
define the so-called Verlinde loop operators \cite{AGGTV,DGOT}. These operators generate a
representation of the quantised algebra of algebraic functions on $\CM_{\rm flat}(C)$ on
the spaces of Virasoro conformal blocks \cite{TV13}. The definition of the Verlinde loop operators
given in \cite{AGGTV,DGOT} can easily be rewritten as deformed traces over products of
the operator-valued monodromy matrices defined in Section \ref{q-mono}.

In the normalisation for the conformal blocks defined by setting $N(\be_2,\al,\be_1)\equiv 1$ in
\rf{normcond} one may analytically continue both the conformal blocks and the corresponding
representation of the Verlinde loop operators with respect to the parameter $c$ to generic complex
values of this parameter. It is not hard to check that
\begin{itemize}
\item the definition of the Verlinde loop operators reduces to taking the {\it ordinary} trace of the
matrices $\SM_k$ defined in Section \ref{proof} at $c=1$,
\item the algebra generated by the Verlinde loop operators becomes {\it commutative}
at this value of the central charge $c$, and
\item the transformation relating Virasoro conformal blocks to tau-functions diagonalizes
all Verlinde loop operators simultaneously with eigenvalues being the trace functions
\rf{classWT}.
\end{itemize}
We note that the quantum counterparts of the coordinates $(\si,\tau)$ that can be defined
away from $c=1$ \cite{TV13} remain non-commutative when $c\ra 1$. However, the algebra
of all operators 
that can be constructed from the quantised coordinates $(\si,\tau)$
contains the important
sub-algebra generated by the Verlinde loop operators. The fact that this sub-algebra becomes
{\it commutative} for $c=1$ leads to the existence of {\it new} representations
for the quantised algebra of functions on $\CM_{\rm flat}(C)$  related to the usual
one by the transformation defined in Section \ref{constr}. This representation is {\it not}
unitarily equivalent to the one studied in \cite{TV13} as the measure defining
the scalar product for $c>25$, the Liouville three-point function, can not be
analytically continued to $c=1$. It should be interesting
to investigate this phenomenon and possible generalisations further.

\subsection{Other relations between isomonodromic deformations and Liouville theory}

There are further relations between
the isomonodromic deformation problem and
Liouville theory: The semiclassical limit
of the null-vector decoupling equations in Liouville theory yields Hamilton-Jacobi - like
equations that define the Hamiltonians generating the isomonodromic
deformation flows.
This was first pointed out in \cite{T11},  a special case was later
rediscovered in \cite{LLNZ}.

It seems remarkable that there exist relations between Liouville
conformal blocks and isomonodromic tau-functions
both in the cases $c=1$ and $c\ra \infty$. A good explanation
remains to be found.

\bigskip

{\bf Acknowledgements.} 
The present work was supported by the Ukrainian SFFR project
F53.2/028, the Program of fundamental research of the physics and
astronomy division of NASU,
project 01-01-14 of NASU, and the IRSES project ``Random and integrable models in
mathematical physics''.

J.T. would like to thank the Euler Institute
(St. Petersburg), where
this work was first presented in the workshop
"Gauge theories and integrability" for hospitality.


\appendix
\section{Calculation of the trace functions}\label{appTF}
Let us compute the trace functions $L_s^r$ and $L_t^r$ in terms of the
parameters $m_{1\ldots 4}^r$, $\sigma_r$, $\tau_r$ using the algorithm developed
in Subsection~\ref{algorithm} along with the rules of Subsection~\ref{Mcalc}. The reader is
referred to Figure~\ref{c04}b (with $p$ replaced by $\sigma_r$) for the labeling of pairs
of pants and boundary components.

The trace functions are determined by the classical monodromies around the punctures $z_1$, $z_2$, $z_3$.
To find them explicitly, we first note that the corresponding operator-valued monodromy matrices are given by
\begin{subequations}
\begin{align}
\SM_1=&\,\big[\SC^{R,0}_{[13]}\SC^{L,1}_{[13]}\cdot\SC\big]^{-1}\big(\SB^R_1\big)^2\,\SC^{R,0}_{[13]}\SC^{L,1}_{[13]}\cdot\SC,\\
\SM_2=&\,\big[\SC^{R,-1}_{[23]}\SC^{L,1}_{[13]}\cdot\SC\big]^{-1}\big(\SB^R_2\big)^2\,\SC^{R,-1}_{[23]}\SC^{L,1}_{[13]}\cdot\SC,\\
\SM_3=&\,\big[\SC^{L,0}_{[23]}\cdot\SC\big]^{-1}\big(\SB^L_2\big)^2\,\SC^{L,0}_{[23]}\cdot\SC.
\end{align}
\end{subequations}
Here the common factor $\SC$ corresponds to the part of analytic continuation path which relates
the base-point $y_0$ to the boundary component $3$ of $C^{L}_{0,3}$ (the neighborhood of the black dot on the
boundary circle in Figure~\ref{pantsdef}). The factor next to it depends on what one wants to achieve at the subsequent step:
the black circle on the boundary $2$ or the empty circle on the boundary~$1$ of $C^{L}_{0,3}$. In the latter
case, for instance, the arc $[13]_L$ should be preceded by the half-turn $b^L_{3}$.

The observations of Subsection~\ref{Mcalc} allow one to get rid of the shift operators
in the computation of classical monodromies by replacing the operator-valued matrices $\SC_{[ji]}^{t,\nu}$ by the ordinary
matrices $C_{[ji]}^{t,\nu}$ defined by (\ref{cnu}). We may therefore set $\SC=1$ in the calculation of the
trace functions. Also note that the resulting expressions are independent of the parameter $\tau_4$ associated to
the boundary curve $\delta_4$: this is a consequence of the factorization
\beq
\left(TB\right)^{t,\nu}_i=\big(\tilde{B}^t_{i}\big)^{-\nu}
\left(\begin{array}{cc}
0 & e^{-\frac{\mathrm i}{2}\tau^t_i} \\
e^{\frac{\mathrm i}{2}\tau^t_i} & 0
\end{array}\right),\qquad \tilde{B}^t_{i}={\mathrm i}\,\sigma_3B^t_i.
\eeq
\nopagebreak
We can now write $L^r_s$, $L^s_t$ as the traces
\begin{subequations}
\begin{align}\notag
 L^r_s=&\,\operatorname{tr}\left(\big[C^{R,-1}_{[23]}\big]^{-1}\big(B^R_2\big)^2\,C^{R,-1}_{[23]}
 \big[C^{R,0}_{[13]}\big]^{-1}\big(B^R_1\big)^2\,C^{R,0}_{[13]}\right)=\\
 =&\,\operatorname{tr}\left( \big(\tilde{B}^R_3\big)^{-1}F^R_{[32]}\big(\tilde{B}^R_2\big)^{2}F^R_{[23]}\,\tilde{B}^R_3 \, F^R_{[31]}
 \big(\tilde{B}^R_1\big)^{2}F^R_{[13]}\right),\\
 \label{traceLt}
  L^r_t=&\, \operatorname{tr}\left(\big[C^{R,-1}_{[23]}C^{L,1}_{[13]}\big]^{-1}\big(B^R_2\big)^2\,C^{R,-1}_{[23]}C^{L,1}_{[13]}\,
 \big[C^{L,0}_{[23]}\big]^{-1}\big(B^L_2\big)^2\,C^{L,0}_{[23]}\right).
\end{align}
\end{subequations}
The first of the equations (\ref{classWT}) then follows from the easily verified identity
\beq
F^t_{[31]} \tilde{B}^t_1 \, F^t_{[12]}\tilde{B}^t_2 \,F_{[23]}^t \tilde{B}^t_3={\mathrm i},
\eeq
which should be understood as a version of the Moore-Seiberg hexagonal relation. To demonstrate the second equation,
observe that (\ref{traceLt}) may be rewritten as
\beqa
&L^r_t= G^R_{+-}G^L_{+-}e^{i\tau_r}+\left(G^R_{++}G^L_{--}+G^R_{--}G^L_{++}\right)+G^R_{-+}G^L_{-+}e^{-i\tau_r},
\notag\\
&G^R=\big[F^R_{[23]}\tilde{B}^R_3\big]^{-1}\big(\tilde{B}^R_2\big)^{2}F^R_{[23]}\tilde{B}^R_3,
\qquad G^L=\tilde{B}_1^L F^L_{[12]}\big(\tilde{B}^L_2\big)^2 \big[\tilde{B}_1^L F^L_{[12]}\big]^{-1}.\notag
\eeqa
The rest of the computation is straightforward.


\end{document}